\documentclass[12pt, onecolumn]{IEEEtran}

\title{Optimizing Orthogonal Multiple Access based on Quantized Channel State Information}

\author{
   {Antonio G. Marques, Georgios B. Giannakis, and Javier Ramos}
   \thanks{Work in this paper was supported by the NSF grants CCF 0830480 and CON 0824007; USDoD ARO Grant No.
W911NF-05-1-0283; C.A. Madrid Grant No. P-TIC-000223-0505; Spanish Government Grant No. TEC2009-12098; and also through collaborative participation in
the Communications and Networks Consortium sponsored by the U. S.
Army Research Laboratory under the Collaborative Technology
Alliance Program, Cooperative Agreement DAAD19-01-2-0011. The U.S. Government is authorized to reproduce and
distribute reprints for Government purposes notwithstanding any copyright
notation thereon. Part
    of this paper was presented in the \emph{IEEE Intl. Workshop on Signal Process. Advances in Wireless Commun.}, Recife, Brasil, July
    2008.}
   \thanks{A. G. Marques and J. Ramos are with the Department of Signal Theory and Communications, Rey Juan Carlos University, Camino
    del Molino s/n, Fuenlabrada, Madrid 28943, Spain, e-mails: antonio.garcia.marques@urjc.es, javier.ramos@urjc.es}
   \thanks{G. B. Giannakis is with the Department of Electrical and Computer Engineering, University of Minnesota, 200
    Union Street SE, Minneapolis, MN 55455, USA, e-mail:
    georgios@umn.edu}}

\usepackage{xspace, amsmath, graphicx, amsfonts, amssymb, syntonly, bm, color, enumerate,bbm}

\include{epsf}
\include{epsfrot}
\include{psfig}
\include{epsfig}

\newcommand{\beq} {\begin{equation}}
\newcommand{\eeq} {\end{equation}}

\newtheorem{lemma}{\hspace{-11pt}\bf Lemma}

\newtheorem{proposition}{\hspace{-11pt}\bf Proposition}
\newtheorem{example}{\hspace{-11pt}\bf Example}

\begin{document}
%
\maketitle
%
\begin{abstract}
The performance of systems where multiple users communicate over wireless fading
links benefits from channel-adaptive allocation of the available
resources. Different from most existing approaches that allocate
resources based on perfect channel state information, this
work optimizes channel scheduling along with per user rate and power
loadings over orthogonal fading channels, when both terminals and scheduler rely
on quantized channel state information. Channel-adaptive policies are designed to
optimize an average transmit-performance criterion subject
to average quality of service requirements. While the resultant optimal policy per
fading realization shows that the individual rate and power loadings
can be obtained separately for each user, the optimal scheduling is
slightly more complicated. Specifically, per fading realization each
channel is allocated either to a single (winner) user, or, to a
small group of winner users whose percentage of shared resources is
found by solving a linear program. A single scheduling scheme
combining both alternatives becomes possible by smoothing the
original disjoint scheme. The smooth scheduling is asymptotically
optimal and incurs reduced computational complexity. Different alternatives to obtain the Lagrange multipliers required to implement the channel-adaptive policies are proposed, including stochastic iterations that are provably convergent and do not require knowledge of the channel distribution. The development
of the optimal channel-adaptive allocation is complemented with discussions on the overhead required to implement the novel policies.
\end{abstract}
\section{Introduction}\label{S:Introduction}

The importance of channel-adaptive allocation of bandwidth, rate, and power
resources in wireless multiuser access over fading links has
been well documented from both information theoretic and practical
communication perspectives \cite{GoldsmithBook}. Per fading realization, parameters
including rate, power and percentages of time frames (or system
subcarriers) are adjusted across users to optimize utility measures
of performance quantified by bit error rate (BER), weighted sum-rate
or power efficiency, under quality of service (QoS) constraints such
as prescribed BER, delay, maximum power or minimum rate
requirements. To carry out such constrained optimization tasks, most
existing approaches assume that perfect CSI (P-CSI) is available
wherever
needed \cite{kn:Tse&Hanly1998}, \cite{kn:Hanly&Tse1998}, \cite{kn:Li&Goldsmith2001a}, \cite{kn:Li&Goldsmith2001b}, \cite{kn:xinGG_it08}, \cite{kn:lata99}.
However, it is well appreciated that errors in estimating the
channel, feedback delay, and the asymmetry between forward and
reverse links render acquisition of deterministically perfect CSI at
transmitters (P-CSIT) impossible in most wireless scenarios
\cite{kn:lash02j}. For cases where the scheduling takes place at the receiver, this has motivated scheduling and resource
allocation schemes using perfect CSI at the receivers (P-CSIR)
but only quantized CSI at the transmitters (Q-CSIT), that can be pragmatically obtained
through finite-rate feedback from the receiver, see, e.g., \cite{kn:amfdgg08}, \cite{kn:xwamgg2008}, and also \cite{SI_jsac_lrf} for a recent review on finite-rate feedback systems.

This work goes one step further to pursue optimal scheduling and
resource allocation for orthogonal multi-access transmissions over
fading links when only Q-CSI is available at the scheduler (as, e.g., \cite{NiharMIMOBroadFRF} for the non-orthogonal multiple input multiple output -MIMO- case),
while transmitters have either perfect or quantized CSI. The
unifying approach minimizes an average power cost (or in a dual
formulation maximizes an average rate utility) subject to
average QoS constraints on rate (respectively power) related
constraints. This setup is particularly suited for systems where the
receiver does not have accurate channel estimates (e.g., when
differential (de-)modulation is employed or when the fading channel
varies fast). It is also pertinent in distributed set-ups (sensor networks or
cellular downlink communications), where the scheduler
(fusion center, access point) is not the receiver and can only acquire
Q-CSI sent by the terminals. The distinct features of this paper
are:
\begin{itemize}
\item Optimal resource allocation schemes that adapt rate, power, and user scheduling as a function of the instantaneous Q-CSI.
\item The optimal rate and power loadings per user terminal depend on the Q-CSI corresponding to its own fading realization, its relative contribution to the power cost (quantified through a user-dependent priority weight), and its rate requirement.
\item The optimal scheduling per channel boils down to one out of two modes: (i) a single user accessing the channel; or, (ii) a small set of users sharing the channel. The channel access coefficients under (ii) are obtained as the solution of a linear program. This bimodal policy emerges not only in systems that operate based on Q-CSI, but also in those that rely on P-CSI but operate over channels whose probability density function (pdf) contains deltas (e.g., discrete random channels or deterministic channels).
\item A novel asymptotically optimum scheduling scheme facilitating convergence and reducing complexity. This scheme combines the aforementioned cases (i) and (ii), and only incurs an $\varepsilon$-loss relative to the optimal solution (with $\varepsilon$ representing a small positive number).
\item Stochastic allocation schemes that are provably convergent, without requiring knowledge of the channel distribution, while reducing the complexity of the overall design.
\item Operating conditions under which the system overhead can be reduced are identified.
\end{itemize}
In addition, the approach here unifies notation at the receiving and transmitting ends, and clarifies the model when Q-CSI is available, yielding valuable insights for improved understanding of channel-adaptive resource allocation and finite-rate feedback.

The rest of the paper is organized as follows. After modeling preliminaries in Section \ref{S:Preliminaries-model},
the general problem is formulated in Section \ref{S:Problem_Formul}, and the optimal solution is
characterized in Section \ref{S:Optimum_Solution}. Algorithms to obtain the optimum Lagrange multipliers needed to implement the optimal policies are developed in Section \ref{S:smooth_and_opt_Lag}. Those algorithms rely on a novel smooth scheduling policy that reduces complexity and guarantees asymptotic optimality. Stochastic scheduling algorithms that do not require knowledge of the channel distribution are also developed. Section \ref{S:Overhead_Red} provides examples and insights on the practical implementation of the novel channel-adaptive schemes.
Numerical tests corroborating the analytical claims are described in Section \ref{S:Simulations}, and concluding remarks are offered in
Section \ref{S:Conclusions}.\footnote{ {\emph Notation:} Boldface upper (lower) case letters
are used for matrix (column vectors); $(\cdot)^T$ denotes transpose;
$[\cdot]_{k,l}$ the $(k,l)$th entry of a matrix, and $[\cdot]_{k}$ the
$(k)$th column (entry) of a matrix (vector); $\odot$ stands for entrywise (Hadamard)
matrix product; $\cdot$ denotes differentiation; $\mathbf{1}$ and
$\mathbf{0}$ are the all-one and all-zero matrices. Calligraphic letters
are used for sets with $|\mathcal{X}|$ denoting cardinality of the set $\mathcal{X}$.
For a random scalar (matrix) variable $x$ ($\mathbf{X}$), the univariate
(multivariate) probability density function (pdf) is denoted by
$f_{x}(x)$ (respectively $f_{\mathbf{X}}(\mathbf{X})$).
Finally, $\wedge$ ($\vee$) denotes the ``and'' (``or'') logic operator,
$x^*$ the optimal value of variable $x$; and, $\mathbbm{1}_{\{\cdot\}}$
the indicator function ($\mathbbm{1}_{\{x\}}=1$ if $x$ is true and
zero otherwise).}\vspace{-0.2cm}

\section{Preliminaries and Problem Statement}\label{S:Preliminaries-model}

Consider a wireless network with $M$ user terminals, indexed by $m\in
\{1,\ldots,M\}$, transmitting over $K$ flat-fading orthogonal
channels, indexed by $k\in \{1,\ldots,K\}$, to a common destination, e.g.,
a fusion center or an access point. Zero-mean additive white Gaussian noise (AWGN) with unit variance is
assumed present at the receiver. With $g_{m,k}$ denoting the $k$th channel's
instantaneous gain (magnitude square of the fading coefficient) between
the $m$th user and the destination, the overall channel is described
by the $M\times K$ gain matrix $\mathbf{G}$ for which $[\mathbf{G}]_{m,k}:=g_{m,k}$.
The range of values each $g_{m,k}$ takes is divided into non-overlapping
regions; and instead of $g_{m,k}$ itself, destination and transmitters
have available only the binary codeword indexing the region $g_{m,k}$
falls into. With $j_{m,k}$ representing the corresponding region index, the
$M\times K$ matrix $\mathbf{J}$ with entries $[\mathbf{J}]_{m,k}:=j_{m,k}$
constitutes the Q-CSI of the overall system. Since $g_{m,k}$ is random,
$j_{m,k}$ is also a discrete random variable; and likewise $\mathbf{J}$ is
random, taking matrix values from a set $\mathcal{J}$ with finite
cardinality $|\mathcal{J}|$.

As in \cite{kn:lata99}, \cite{kn:amfdgg08}, \cite{kn:Li&Goldsmith2001a} or \cite{kn:xinGG_it08}, users at the outset can be
scheduled to access simultaneously but orthogonally (in time or frequency)
any of the $K$ channels. The channel scheduling policy is described by an
$M\times K$ matrix $\mathbf{W}$ whose nonnegative entry $[\mathbf{W}]_{m,k}$ corresponds
to the \emph{percentage} of the $k$th channel scheduled for the $m$th user. Clearly,
it holds that $\sum_{m=1}^M[\mathbf{W}]_{m,k} \in [0,1]$ $\forall k$. The
power and rate resources of all terminal-channel pairs are collected in $M\times K$ matrices ${\mathbf{P}}$
and ${\mathbf{R}}$, respectively. Each of the corresponding entries $[\mathbf{P}]_{m,k}$ and
$[\mathbf{R}]_{m,k}$ represent, respectively, the {\it nominal} power and rate
the $m$th user terminal would be allocated if it were the only terminal
scheduled to transmit over the $k$th channel. Note that such entries are lower bounded by zero and upper bounded by the maximum nominal power and rate that the hardware of the system is able to implement. Since scheduling and allocation
will be adapted based on Q-CSI, matrices $\mathbf{W}$,
$\mathbf{P}$ and $\mathbf{R}$ will depend on $\mathbf{J}$ and each
can take at most $|\mathcal{J}|$ different values. Under prescribed BER or capacity
constraints, rate and power variables are coupled. This power-rate coupling
will be represented by a function $\Upsilon$ (respectively $\Upsilon^{-1}$
for the rate-power coupling), which relates $[{\mathbf{P}}]_{m,k}$ to
$[{\mathbf{R}}]_{m,k}$ over the same Q-CSI region $\mathcal{R}([\mathbf{J}]_{m,k})$. (Wherever needed, we will
write $\Upsilon_{\mathcal{R}([\mathbf{J}]_{m,k})}$ to exemplify this
dependence.)

\subsection{Problem Formulation}\label{S:Problem_Formul}

Given the Q-CSI matrix $\mathbf{J}$ and prescribed QoS requirements,
the goal is to find $\mathbf{W(J)}$, ${\mathbf{P(J)}}$ and
${\mathbf{R(J)}}$ so that the \emph{overall} average \emph{weighted}
performance is optimized. (Overall here refers to performance of all
users and weighted refers to different user priorities effected
through a preselected weight vector $\boldsymbol{\mu}:=[\mu_1,\ldots,$
$\mu_M]^T$ with nonnegative entries.) Depending on desirable
objectives, the problem can be formulated either as constrained
utility maximization of the average weighted sum-rate subject to
average power constraints; or, as a constrained minimization of the
average weighted power subject to average rate constraints.
The former fits the classical rate
(capacity) maximization, while the latter is particularly relevant in
energy-limited scenarios (e.g., sensor networks) where power
savings is the main objective.
Although this paper will use the power minimization formulation, the rate maximization problem
can be tackled readily by dual substitutions;
namely, after interchanging the roles of ${\mathbf{R}}$ and
$\Upsilon_{\mathcal{R}([\mathbf{J}]_{m,k})}$ by ${\mathbf{P}}$
and $\Upsilon_{\mathcal{R}([\mathbf{J}]_{m,k})}^{-1}$, respectively.

Specifically, the weighted average transmit-power will be minimized
subject to individual minimum average rate constraints collected
in the vector $\mathbf{\check{r}}:=[\check{r}_{1},\ldots,\check{r}_{M}]^T$.
Per Q-CSI realization $\mathbf{J}$, the overall weighted transmit-power is given by
$\sum_{m=1}^{M}[\boldsymbol{\mu}]_m\sum_{k=1}^{K}[{\mathbf{P(\mathbf{J})}}]_{m,k}
[\mathbf{W(\mathbf{J})}]_{m,k}$; while the $m$th user's transmit-rate is
$\sum_{k=1}^{K}[{\mathbf{R(\mathbf{J})}}]_{m,k} [\mathbf{W(\mathbf{J})}]_{m,k}$.
Using the probability mass function $\Pr\{\mathbf{J}\}$, these expressions
can be used to obtain the average transmit-power and transmit-rate. For a
given channel quantizer, i.e., with $\mathcal{R}$ fixed, and the fading pdf
assumed known, $\Pr\{\mathbf{J}\}$ can be obtained as $\Pr\{\mathbf{J}\}=
\int_{\mathcal{R}(\mathbf{J})} f_{\mathbf{G}}(\mathbf{G})d\mathbf{G}$, where $\mathcal{R}(\mathbf{J})$ represents the region of the $\mathbf{G}$ domain such that $\mathbf{G}\in \mathcal{R}(\mathbf{J})$ are quantized as $\mathbf{J}$.
Since $\Upsilon_{\mathcal{R}([\mathbf{J}]_{m,k})}$ links ${\mathbf{R}}$
with ${\mathbf{P}}$, it suffices to optimize only over one of them. Note
also that the binomial $[\mathbf{{R}(J)}]_{m,k}[\mathbf{W(\mathbf{J})}]_{m,k}$
is not jointly convex with respect to (w.r.t.)  $\mathbf{{R}(J)}$ and
$\mathbf{{W}(J)}$. For this reason, we will instead consider the auxiliary variable
$[\mathbf{\tilde{R}(J)}]_{m,k}:=[\mathbf{{R}(J)}]_{m,k}[\mathbf{W(\mathbf{J})}]_{m,k}$
and seek allocation and scheduling matrices solving the following optimization problem:
\begin{equation}\label{E:optRA}
\begin{cases}
        \min_{\mathbf{\tilde{R}(J)\geq0,W(J)\geq0}}
       {\displaystyle \sum_{\forall \mathbf{J} \in \mathcal{J}}}\left(\sum_{m=1}^{M}[\boldsymbol{\mu}]_m\sum_{k=1}^{K}\right.
       \left.\Upsilon_{\mathcal{R}([\mathbf{J}]_{m,k})}\left(\frac{[{\mathbf{\tilde{R}(\mathbf{J})}}]_{m,k}}{[\mathbf{W(\mathbf{J})}]_{m,k}}\right)[\mathbf{W(\mathbf{J})}]_{m,k}\right)\Pr\{\mathbf{J}\} \\
        \mathrm{s.~to:} \;\; {\displaystyle \sum_{\forall \mathbf{J} \in \mathcal{J}}}\left(\sum_{k=1}^{K} [{\mathbf{\tilde{R}(\mathbf{J})}}]_{m,k}\right)\Pr\{\mathbf{J}\}\geq [\mathbf{\check{r}}]_m, \;\; \forall
              m
        \\
        \hspace{1.1cm}  \sum_{m=1}^M [\mathbf{W(\mathbf{J})}]_{m,k} \leq 1, \;\; \forall
        k,\forall\mathbf{J}\;.
\end{cases}
\end{equation}

Appendix A shows that if $\Upsilon_{\mathcal{R}([\mathbf{J}]_{m,k})}$ is a convex function, then
problem \eqref{E:optRA} is convex. Throughout this paper it will be assumed that:\\
{\bf (as1)} {\it the power-rate function $\Upsilon_{\mathcal{R}([\mathbf{J}]_{m,k})}$ is increasing and strictly
convex}.\\
This assumption holds generally true for orthogonal access but, for example, not
when multiuser interference is present. Note also that (as1) implies that the rate-power function $\Upsilon^{-1}$ is increasing and strictly concave. To justify the adoption of (as1), consider the following example of $\Upsilon$.
\begin{example}\label{Fot:footnote_BER}
For simplicity, the tractable case of outage capacity will be consider here, postponing the case of ergodic capacity to Section \ref{S:examplesofUpsilon}. Suppose that we want the outage probability of the $m$th user over the $k$th channel for a given Q-CSI $\mathbf{J}$ to be $\delta$. Define the $\delta$-outage channel gain for the $(m,k)$ pair in $\mathcal{R}([\mathbf{J}]_{m,k})$ as $g_{m,k}^{\delta}([\mathbf{J}]_{m,k})$ so that $\Pr\{g_{m,k}\leq g_{m,k}^{\delta}([\mathbf{J}]_{m,k})~|~g_{m,k}\in\mathcal{R}([\mathbf{J}]_{m,k})\}=\delta$. Then using Shannon's capacity formula, the rate-power function can be written as $\Upsilon_{\mathcal{R}([\mathbf{J}]_{m,k})}^{-1}(x)=\log_2(1+xg_{m,k}^{\delta}([\mathbf{J}]_{m,k}))$. Solving the previous expression w.r.t. $x$, yields the power-rate function $\Upsilon_{\mathcal{R}([\mathbf{J}]_{m,k})}\left(x\right)=(2^{x}-1)/g_{m,k}^{\delta}([\mathbf{J}]_{m,k})$, which is certainly increasing and strictly convex as required by (as1).
\end{example}

Before moving to the next section where the solution of \eqref{E:optRA} will be characterized, it is important to stress that since $\mathcal{R}$ is involved in specifying
$\Pr\{\mathbf{J}\}$ and
$\Upsilon_{\mathcal{R}([\mathbf{J}]_{m,k})}$, the choice of
$\mathcal{R}$ affects the optimum allocation. Selecting the
quantization regions to optimize (\ref{E:optRA}) is thus of interest
but goes beyond the scope of this paper. Near-optimal channel
quantizers for time division multiple access (TDMA) and orthogonal frequency-division multiple access (OFDMA)
can be found in \cite{kn:xwamgg2008} and \cite{kn:amfdgg08}, respectively.

\section{Optimum Resource Allocation}\label{S:Optimum_Solution}

In this section, the optimum $\mathbf{W}$, $\mathbf{P}$ and $\mathbf{R}$
matrices will be characterized as a function of $\mathbf{J}$ and the
optimum multipliers of the constrained optimization problem in \eqref{E:optRA}.

Let $\boldsymbol{\lambda}^{R}$ denote the $M\times 1$ vector whose entries are the non-negative Lagrange multipliers associated with the $m$th
\emph{average} rate constraint; and $\boldsymbol{\lambda}^{W}(\mathbf{J})$ the $K\times 1$ vector corresponding to the $k$th
channel-sharing constraint \emph{per Q-CSI} matrix\footnote{The dependence of the multipliers associated with instantaneous constraints on $\mathbf{J}$ will be explicitly written throughout.}  $\mathbf{J}$. Let also
$\boldsymbol{\alpha}^{R}(\mathbf{J})$ and $\boldsymbol{\alpha}^{W}(\mathbf{J})$
denote $K \times M$ matrices whose entries are, correspondingly, the non-negative
Lagrange multipliers associated with the constraints
$[\mathbf{\tilde{R}(J)}]_{m,k}\geq0$ and
$[\mathbf{W(J)}]_{m,k}\geq0$. The full Lagrangian of (\ref{E:optRA}) can be written as
\begin{eqnarray}\label{E:Lagrangian}
\nonumber\mathcal{L}(\boldsymbol{\lambda}^R,\boldsymbol{\lambda}^{W}(\mathbf{J}),\boldsymbol{\alpha}^{R}(\mathbf{J}),
\boldsymbol{\alpha}^{W}(\mathbf{J}),\mathbf{\tilde{R}}(\mathbf{J}),\mathbf{W}(\mathbf{J})):=\\
\nonumber \sum_{\forall \mathbf{J} \in \mathcal{J}}\left(\sum_{m=1}^{M}[\boldsymbol{\mu}]_m\sum_{k=1}^{K}\Upsilon_{\mathcal{R}
([\mathbf{J}]_{m,k})}\left(\frac{[{\mathbf{\tilde{R}(\mathbf{J})}}]_{m,k}}{[\mathbf{W(\mathbf{J})}]_{m,k}}\right)
[\mathbf{W(\mathbf{J})}]_{m,k}\right)\Pr\{\mathbf{J}\}\\
\nonumber- \sum_{m=1}^M \left( [\boldsymbol{\lambda}^R]_m \sum_{\forall \mathbf{J} \in \mathcal{J}}\left(\sum_{k=1}^{K} [{\mathbf{\tilde{R}(\mathbf{J})}}]_{m,k}\right)\Pr\{\mathbf{J}\} - [\mathbf{\check{r}}]_m\right)
+\sum_{\forall \mathbf{J} \in \mathcal{J}}\sum_{k=1}^{K}
[\boldsymbol{\lambda}^{W}(\mathbf{J})]_{k}\left(\sum_{m=1}^{M}[\mathbf{W(\mathbf{J})}]_{m,k}-1\right)\\
-\sum_{\forall \mathbf{J} \in \mathcal{J}}\sum_{m=1}^{M}\sum_{k=1}^{K}
\left([\boldsymbol{\alpha}^{R}(\mathbf{J})]_{m,k}[\mathbf{\tilde{R}(J)}]_{m,k}
+[\boldsymbol{\alpha}^{W}(\mathbf{J})]_{m,k}[\mathbf{W}(\mathbf{J})]_{m,k}\right).
\end{eqnarray}
Because (\ref{E:optRA}) is convex, the Karush-Kuhn-Tucker (KKT)
conditions yield the following necessary and
sufficient conditions of optimality \cite{kn:Bertsekas99} (recall $\dot{x}$ denotes the derivative of $x$):
\begin{eqnarray}
\label{E:kktr}\hspace{2.1cm}[\boldsymbol{\mu}]_m\dot{\Upsilon}_{\mathcal{R}([\mathbf{J}]_{m,k})}\left(\frac{[\mathbf{\tilde{R}^{*}(J)}]_{m,k}}
{[\mathbf{{W}^{*}(J)}]_{m,k}}\right)\Pr\{\mathbf{J}\}-[\boldsymbol{\lambda}^{R*}(\mathbf{J})]_m\Pr\{\mathbf{J}\}
-[\boldsymbol{\alpha}^{R*}(\mathbf{J})]_{m,k}&=&0\\
\label{E:kktrok}[\mathbf{\tilde{R}^{*}(J)}]_{m,k}[\boldsymbol{\alpha}^{R*}(\mathbf{J})]_{m,k}&=&0
\end{eqnarray}
\begin{eqnarray}
\nonumber\hspace{-0.8cm}[\boldsymbol{\mu}]_m\Upsilon_{\mathcal{R}([\mathbf{J}]_{m,k})}\left(\frac{[\mathbf{\tilde{R}^{*}(J)}]_{m,k}}{[\mathbf{{W}^{*}(J)}]_{m,k}}\right)\Pr\{\mathbf{J}\}-
[\boldsymbol{\mu}]_m\dot{\Upsilon}_{\mathcal{R}([\mathbf{J}]_{m,k})}\left(\frac{[\mathbf{\tilde{R}^{*}(J)}]_{m,k}}{[\mathbf{{W}^{*}(J)}]_{m,k}}\right)
\frac{[\mathbf{\tilde{R}^{*}(J)}]_{m,k}}{[\mathbf{{W}^{*}(J)}]_{m,k}}\Pr\{\mathbf{J}\}&~&\\
\label{E:kktt}-[\boldsymbol{\alpha}^{W*}(\mathbf{J})]_{m,k}+[\boldsymbol{\lambda}^{W*}(\mathbf{J})]_{k}&=&0\\
\label{E:kkttok}[\mathbf{W^{*}(J)}]_{m,k}[\boldsymbol{\alpha}^{W*}(\mathbf{J})]_{m,k}&=&0.
\end{eqnarray}
Conditions (\ref{E:kktr})-(\ref{E:kkttok}) can be used to characterize the optimal rate and channel allocation as follows.
\begin{proposition}\label{P:propRAll}
{\em The optimum rate allocation is given by: \\
(i) $[\mathbf{\tilde{R}^{*}(J)}]_{m,k}=0$, if either
$[\mathbf{{W}^{*}(J)}]_{m,k}=0$ or $[\boldsymbol{\lambda}^{R*}]_m/[\boldsymbol{\mu}]_m<\dot{\Upsilon}_{\mathcal{R}([\mathbf{J}]_{m,k})}$ $\left(\frac{[\mathbf{\tilde{R}^{*}(J)}]_{m,k}}{[\mathbf{{W}^{*}(J)}]_{m,k}}\right)$; otherwise,
\\
(ii) the optimum rate allocation is
\begin{equation}\label{E:rateopt_ef}
[\mathbf{\tilde{R}^{*}(J)}]_{m,k}=\dot{\Upsilon}_{\mathcal{R}([\mathbf{J}]_{m,k})}^{-1}
\left(\frac{[\boldsymbol{\lambda}^{R*}]_{m}}{[\boldsymbol{\mu}]_m}\right)[\mathbf{{W}^{*}(J)}]_{m,k}
\end{equation}
where $\dot{\Upsilon}_{\mathcal{R}([\mathbf{J}]_{m,k})}^{-1}$ denotes the
inverse function of
$\dot{\Upsilon}_{\mathcal{R}([\mathbf{J}]_{m,k})}$.}
\end{proposition}
\begin{IEEEproof}
Consider first the claim in (i). The definition of $[\mathbf{\tilde{R}^{*}(J)}]_{m,k}$ implies that if $[\mathbf{{W}^{*}(J)}]_{m,k}=0$, then $[\mathbf{\tilde{R}^{*}(J)}]_{m,k}=0$. On the other hand, if $[\boldsymbol{\lambda}^{R*}]_m/[\boldsymbol{\mu}]_m<\dot{\Upsilon}_{\mathcal{R}([\mathbf{J}]_{m,k})}(\cdot)$, then \eqref{E:kktr} can only be satisfied if $[\boldsymbol{\alpha}^{R*}(\mathbf{J})]_{m,k}>0$. Using the slackness condition in \eqref{E:kktrok}, the latter implies $[\mathbf{\tilde{R}^{*}(J)}]_{m,k}=0$.
The proof of part (ii) is simpler and consists of solving \eqref{E:kktr} after excluding the two cases in (i); i.e. assuming that $[\mathbf{{W}^{*}(J)}]_{m,k}>0$ and $[\boldsymbol{\alpha}^{R*}(\mathbf{J})]_{m,k}=0$.
\end{IEEEproof}
Given the relationship between $\mathbf{\tilde{R}}$ and $\mathbf{R}$,
the optimum transmit-rate for $[\mathbf{{W}^{*}(J)}]_{m,k}\neq0$ is
\begin{equation}\label{E:rateopt_nom}
[\mathbf{R^{*}(J)}]_{m,k}=\dot{\Upsilon}_{\mathcal{R}([\mathbf{J}]_{m,k})}^{-1}
\left(\frac{[\boldsymbol{\lambda}^{R*}]_{m}}{[\boldsymbol{\mu}]_m}\right).
\end{equation}
In fact, \eqref{E:rateopt_nom} is also valid if $[\mathbf{{W}^{*}(J)}]_{m,k}=0$. This is because when $[\mathbf{{W}^{*}(J)}]_{m,k}=0$, any finite nominal rate yields $[\mathbf{R^{*}(J)}]_{m,k}=0$, which is the optimal solution.
Equation \eqref{E:rateopt_nom} shows that the optimal rate loading depends on the ratio of $[\boldsymbol{\mu}]_m$ over $[\boldsymbol{\lambda}^{R*}]_m$, where the first represents the ``priority'' terminal $m$ has to minimize the total power cost, and the latter represents the price corresponding its rate requirement. According to (as1), $\dot{\Upsilon}$ is monotonically increasing function and so is $\dot{\Upsilon}^{-1}$ in \eqref{E:rateopt_nom}. This implies that users with high $[\check{\mathbf{r}}]_m$ have high values of $[\boldsymbol{\lambda}^{R*}]_m$, thus higher rate and power loadings per region. Conversely, for users whose power consumption is critical the optimum solution sets high values of $[\boldsymbol{\mu}]_m$, thus low rate and power loadings per region. Part (i) of the proposition also dictates that there may be regions for which the optimum rate and power loadings are zero. Intuitively, this will typically happen for the region(s) whose channel conditions are so poor that the power cost of activating the region may be too high.

To find the optimum scheduling matrix $\mathbf{W}$, define first the functional
\begin{eqnarray}\label{E:phi_chann_ind}
[\mathbf{C}_W(\mathbf{J})]_{m,k}&:=&[\boldsymbol{\mu}]_m\Upsilon_{\mathcal{R}([\mathbf{J}]_{m,k})}([\mathbf{R^{*}(J)}]_{m,k})-[\boldsymbol{\lambda}^{R*}]_m[\mathbf{R^{*}(J)}]_{m,k}
\end{eqnarray}
which represents the cost of scheduling channel $k$ to user $m$ when the
Q-CSI is $\mathbf{J}$. This cost of selecting $[\mathbf{W(J)}]_{m,k}=1$ emerges also in the two first terms of $\mathcal{L}$ in (\ref{E:Lagrangian}). Based on \eqref{E:phi_chann_ind}, and with $\wedge$ denoting the ``and'' operator, we define the $K\times1$ vector $\mathbf{c}_W^*(\mathbf{J},\boldsymbol{\lambda}^R)$ with entries $[\mathbf{c}_W^*(\mathbf{J},\boldsymbol{\lambda}^R)]_k:=\min_m\{[\mathbf{C}_W$
$(\mathbf{J},\boldsymbol{\lambda}^R)]_{m,k}\}_{m=1}^M$, and the sets of ``winner user(s)''
$\mathcal{M}(\mathbf{J},k):=\{m:
[\mathbf{C}_W(\mathbf{J},\boldsymbol{\lambda}^R)]_{m,k}=[\mathbf{c}_W^*(\mathbf{J},$
$\boldsymbol{\lambda}^R)]_k~ \wedge
~([\mathbf{c}_W^*(\mathbf{J},\boldsymbol{\lambda}^R)]_k<0)\}$. Given the Q-CSI realization $\mathbf{J}$, $\mathcal{M}(\mathbf{J},k)$ is the set of user(s) that incur the minimum cost if scheduled to access channel $k$ while $[\mathbf{c}_W^*(\mathbf{J},\boldsymbol{\lambda}^R)]_k$ is the cost corresponding to those users. Using these
notational conventions, it can be shown that:
\begin{proposition}\label{P:propChAll}
{\em The optimum scheduling $\mathbf{W^{*}(J)}$ satisfies the following:\\
(i) If $[\mathbf{W^{*}(J)}]_{m,k}>0$, then $m\in\mathcal{M}(\mathbf{J},k)$; \\
(ii) If $|\mathcal{M}(\mathbf{J},k)|>0$, then $\sum_{m\in
\mathcal{M}(\mathbf{J},k)}[\mathbf{W^{*}(J)}]_{m,k}=1$; and \\
(iii) If $|\mathcal{M}(\mathbf{J},k)|=0$, then
$[\mathbf{W^{*}(J)}]_{m,k}=0$ $\forall m$.}
\end{proposition}
\begin{IEEEproof}
Appendix B.
\end{IEEEproof}

In words, the optimal scheduler assigns the channel only to user(s) with
minimum negative cost \eqref{E:phi_chann_ind}, which is in most cases (but not all) attained by a single user. This is a greedy policy because only one user with minimum cost is selected
to transmit per Q-CSI realization, while others defer. Note that with P-CSIR,
the optimum scheduling over orthogonal fading channels is also greedy, whether
based on P-CSIT \cite{kn:Li&Goldsmith2001a}, \cite{kn:xinGG_it08} or Q-CSIT \cite{kn:amfdgg08}.

\noindent \emph{Case 1 (Single winner user):} When the minimum cost is attained by
only one user, $\mathbf{{W}^{*}}$ in Proposition 2 can be written using the
indicator function, as
\begin{eqnarray}\label{E:simpl_WW}
[\mathbf{{W}^{*}}(\mathbf{J})]_{m,k} =
\mathbbm{1}_{\{m\in\mathcal{M}(\mathbf{J},k)\}}\;.
\end{eqnarray}
Since $[\mathbf{C}_W(\mathbf{J})]_{m,k}$ is a function of different variables (namely, the quantization regions, the
fading realization, the individual priority weight and the individual Lagrange
multiplier), for most CSI realizations the costs corresponding to
different users $m$ are distinct, and the emerging winner is unique.

\noindent \emph{Case 2 (Multiple winners):} The event of having different
users attaining the minimum cost will be henceforth referred to as a
``tie''. The main difficulty with a tie is that Proposition 2-\emph{(ii)}
does not specify how the channel should be split among winner
users (the underlying reason being that any arbitrary allocation minimizes $\mathcal{L}$). On
the other hand, only a subset (for most realizations one) of them
is the actual solution to the original primal problem. To find the
optimum schedule in this case, define first the matrix of single-winner scheduling as
$[\mathbf{W}_{one}(\mathbf{J})]_{m,k}:=[\mathbf{W}^*(\mathbf{J})]_{m,k}$ in
(\ref{E:simpl_WW}) for all $(\mathbf{J},k)$ so that
$|\mathcal{M}(\mathbf{J},k)|=1$, and
$[\mathbf{W}_{one}(\mathbf{J})]_{m,k}:=0$, otherwise. Define further the scheduling matrix with multiple winners as $[\mathbf{W}_{tie}(\mathbf{J})]_{m,k}=0$
if $|\mathcal{M}(\mathbf{J},k)|\leq1$ or if $|\mathcal{M}(\mathbf{J},k)|>1$ but $m\notin \mathcal{M}(\mathbf{J},k)$, and $[\mathbf{W}_{tie}(\mathbf{J})]_{m,k}\in[0,1]$, otherwise. And finally, let the set of multiple-winner scheduling matrices be $\mathcal{W}_{tie}:=\{\mathbf{W}_{tie}(\mathbf{J})~|~\forall \mathbf{J}\}$; the average
single-winner transmit-rate vector $[\mathbf{\bar{r}}_{one}]_m:=$
$\sum_{\forall \mathbf{J}}\left(\sum_{k=1}^{K}[{\mathbf{R^*(\mathbf{J})}}]_{m,k}[\mathbf{W}_{one}(\mathbf{J})]_{m,k}\right)
\Pr\{\mathbf{J}\}$; and
$\mathbf{\check{r}}_{tie}:=\mathbf{\check{r}}-\mathbf{\bar{r}}_{one}$.
Using these definitions, the optimum schedule $\mathbf{W}_{tie}(\mathbf{J})$ for all $(\mathbf{J},k)$ with
$|\mathcal{M}(\mathbf{J},k)|>1$, can be found as the solution of
the following linear program: \vspace{-0.03cm}
\begin{equation}\label{E:optW_tie}
\begin{cases}
        \min_{\mathbf{W}_{tie}(\mathbf{J})\in \mathcal{W}_{tie}}
       \hspace{0.3cm}\sum_{\forall \mathbf{J}}\left(\sum_{k=1}^{K}\sum_{m=1}^M[\boldsymbol{\mu}]_m\right.\left.\Upsilon_{\mathcal{R}([\mathbf{J}]_{m,k})}\left([{\mathbf{R^*(\mathbf{J})}}]_{m,k}\right)[\mathbf{W}_{tie}(\mathbf{J})]_{m,k}\right)\Pr\{\mathbf{J}\} \vspace{0.2cm}\\
        \mathrm{s.~to:} \;\; \sum_{\forall \mathbf{J}}\left(\sum_{k=1}^{K} [{\mathbf{R^*(\mathbf{J})}}]_{m,k}[\mathbf{W}_{tie}(\mathbf{J})]_{m,k}\right)\Pr\{\mathbf{J}\} = [\mathbf{\check{r}}_{tie}]_m, \;\; \forall
              m
        \vspace{0.2cm}\\
        \hspace{1.2cm}  \sum_{m=1}^M [\mathbf{W}_{tie}(\mathbf{J})]_{m,k} = 1, \;\; \forall
        (\mathbf{J},k):~|\mathcal{M}(\mathbf{J},k)|>1.
\end{cases}
\end{equation}
Note that in the optimization process, only the matrices $\mathbf{J}$ for which a tie occurs are considered and for those only the non-zero entries of $\mathbf{W}_{tie}(\mathbf{J})$ are optimized.

The main idea behind (\ref{E:optW_tie}) is that among all schedules minimizing the Lagrangian when a tie occurs (second
constraint), the optimal one for the primal problem is the one
for which the average rate constraints are satisfied with equality. We stress that here $\mathbf{R}^*(\mathbf{J})$ (thus $\mathbf{P}^*(\mathbf{J})$) are fixed and therefore only optimization over the channel-sharing coefficients for which a tie occurs (which in general is a small set) is carried out. To clarify this point, let us consider the following example.
\begin{example}\label{Ex:SystFewStates}
Consider a system with $K=1$ channel, $M=4$ users and $10$ regions per user. For such a system, the number of channel realizations is $|\mathcal{J}|=10^4$. Among those it is found that, e.g., ties occur for 3 different fading realizations, namely: when $\mathbf{J}=\mathbf{J}_1$ users 1 and 2 tie; when $\mathbf{J}=\mathbf{J}_2$ users 1, 3 and 4 tie; and when $\mathbf{J}=\mathbf{J}_3$ users 2 and 4 tie. In this case, the optimization in (\ref{E:optW_tie}) has to be carried out over $[\mathbf{W}(\mathbf{J}_1)]_{1,1}$, $[\mathbf{W}(\mathbf{J}_1)]_{2,1}$, $[\mathbf{W}(\mathbf{J}_2)]_{1,1}$, $[\mathbf{W}(\mathbf{J}_2)]_{3,1}$, $[\mathbf{W}(\mathbf{J}_2)]_{4,1}$, $[\mathbf{W}(\mathbf{J}_3)]_{2,1}$, and $[\mathbf{W}(\mathbf{J}_3)]_{4,1}$. Once $\mathbf{W}_{tie}^*(\mathbf{J})$ is found, the overall optimal channel assignment is $[\mathbf{W(J)}^*]_{m,k}:=[\mathbf{W}_{one}^*(\mathbf{J})]_{m,k}$ for $(\mathbf{J},k)$ with $|\mathcal{M}(\mathbf{J},k)|\leq1$ and $[\mathbf{W}^*(\mathbf{J})]_{m,k}:=[\mathbf{W}_{tie}^*(\mathbf{J})]_{m,k}$ otherwise.
\end{example}

It is worth noticing that for every scenario where multiple users access the channel orthogonally, the optimum scheduling needs to satisfy (\ref{E:optW_tie}). However, neither
\cite{kn:Li&Goldsmith2001a}, \cite{kn:xinGG_it08} (P-CSIR and P-CSIT)
nor \cite{kn:amfdgg08}, \cite{kn:xwamgg2008} (P-CSIR and Q-CSIT) consider (\ref{E:optW_tie}).
This is because if the fading distributions are continuous and P-CSIR is available, the set of fading realizations
$\mathbf{G}$ for which a tie occurs has Lebesgue measure zero. Therefore,
any arbitrary channel scheduling among tied users is equally optimum.
Indeed, the contribution of any specific $\mathbf{G}$ to the average
performance when integrated over the channel pdf is zero. But when
dealing with Q-CSI (or with deterministic fixed channels), neither the probability of a Q-CSI realization
$\mathbf{J}$ nor the contribution to the average cost are negligible.
And this precisely necessitates solving (\ref{E:optW_tie}) to obtain
the optimum schedule. Intuitively, as the number of regions and channels
increases sharing a channel becomes less likely, which in turn brings the
solution closer to the continuous fading P-CSIR case and the effect of neglecting
(\ref{E:optW_tie}) becomes less harmful. The opposite behavior arises in systems that have P-CSIR but further operate over deterministic (fixed) channels. In those systems ties will represent the prevailing channel allocation (e.g., for a deterministic TDMA system we have $K=1$ and $|\mathcal{J}|=1$; since all the users have to access the channel to satisfy their rate constraints, the entries of $\boldsymbol{\lambda}^{R*}$ will self-adjust so that \emph{a tie among all the users} occurs). Only in systems operating over deterministic channels for which the number of channels is much higher than the number of users (e.g., an OFDMA system with many subcarriers), the single-winner case will constitute the predominant scheduling.

In the context of smooth optimization, a single scheduling scheme that can be implemented both for cases 1 and 2, is asymptotically optimal, incurs reduced computational burden and facilitates computation of the optimal Lagrange multipliers is developed in the next section.

\section{Optimal Lagrange Multipliers}\label{S:smooth_and_opt_Lag}

To implement the optimum scheduling and rate allocation policies presented in the previous section, the optimum multiplier vector $\boldsymbol{\lambda}^{R*}$ needs to be known. Since
the rate constraints in (\ref{E:optRA}) are always active, the KKT
conditions imply that when $\boldsymbol{\lambda}^R=\boldsymbol{\lambda}^{R*}$
those constraints are satisfied with equality. Since $\boldsymbol{\lambda}^{R*}$
cannot be obtained analytically from this condition, numerical search is required. This is possible using dual methods. First, let us write\footnote{Throughout this section,
dependence on $\boldsymbol{\lambda}^R$ will be made explicit wherever it
contributes to clarity.} a simplified version of the Lagrangian\vspace{-0.2cm}
\begin{eqnarray}\label{E:simpl_Lagrangian}
\nonumber\mathcal{L}(\boldsymbol{\lambda}^R,\mathbf{\tilde{R}}(\mathbf{J}),\mathbf{W}(\mathbf{J})):=
\sum_{\forall \mathbf{J} \in \mathcal{J}}\left(\sum_{m=1}^{M}[\boldsymbol{\mu}]_m\sum_{k=1}^{K}\Upsilon_{\mathcal{R}
([\mathbf{J}]_{m,k})}\left(\frac{[{\mathbf{\tilde{R}(\mathbf{J})}}]_{m,k}}{[\mathbf{W(\mathbf{J})}]_{m,k}}\right)
[\mathbf{W(\mathbf{J})}]_{m,k}\right)\Pr\{\mathbf{J}\}\\
- \sum_{m=1}^M \left( [\boldsymbol{\lambda}^R]_m \sum_{\forall \mathbf{J} \in \mathcal{J}}\left(\sum_{k=1}^{K} [{\mathbf{\tilde{R}(\mathbf{J})}}]_{m,k}\right)\Pr\{\mathbf{J}\}\right) +\sum_{m=1}^M [\boldsymbol{\lambda}^R]_m[\mathbf{\check{r}}]_m
\end{eqnarray}
where only the contribution of the average rate constraints is considered  [cf. (\ref{E:Lagrangian})]. Because all the instantaneous constraints (i.e., channel-sharing and non-negativity constraints) were already satisfied when obtaining the solution of the previous section, the focus here is to find $\boldsymbol{\lambda}^R$ so that the average rate constraints are satisfied. Let $\mathcal{F}(\mathbf{J})$ denote the feasible set of the rate and channel assignment matrices, namely $\mathcal{F}(\mathbf{J}):=\{(\tilde{\mathbf{R}}(\mathbf{J}),\mathbf{W}(\mathbf{J}))~|~ \mathbf{\tilde{R}}(\mathbf{J})\geq
\mathbf{0} ~ \wedge ~ \mathbf{W}(\mathbf{J})\geq
\mathbf{0} ~ \wedge ~ \sum_{m=1}^M[\mathbf{W}(\mathbf{J})]_{m,k}\leq
1\}$. The dual function is then defined as
\begin{eqnarray}\label{E:dual}
\nonumber \hspace{-0.3cm}D(\boldsymbol{\lambda}^R)
\hspace{-0.3cm}&:=&\hspace{-0.3cm}\inf_{(\mathbf{\tilde{R}}(\mathbf{J}),
\mathbf{W}(\mathbf{J}))\in \mathcal{F}(\mathbf{J})}\mathcal{L}(\boldsymbol{\lambda}^R,\mathbf{\tilde{R}}(\mathbf{J}),\mathbf{W}(\mathbf{J}))\\
&=&\hspace{-0.3cm}\mathcal{L}(\boldsymbol{\lambda}^R,\mathbf{R}^*(\mathbf{J},\boldsymbol{\lambda}^R)\odot\mathbf{W}^*(\mathbf{J},\boldsymbol{\lambda}^R),\mathbf{W}^*(\mathbf{J},\boldsymbol{\lambda}^R))
\end{eqnarray}
which is concave w.r.t. $\boldsymbol{\lambda}^R$. Based on (\ref{E:dual}), the dual problem of (\ref{E:optRA}) is
\begin{equation}\label{E:dual_problem}
\max_{\boldsymbol{\lambda}^R\geq \mathbf{0}}
D(\boldsymbol{\lambda}^R).
\end{equation}
Since the problem in \eqref{E:optRA} is convex and strictly feasible, the duality gap between the primal and dual problems is zero. Thus, the value of
$\boldsymbol{\lambda}^R$ optimizing (\ref{E:dual_problem}) can
be used to find the optimum primal solution.
A standard approach to obtain $\boldsymbol{\lambda}^{R*}$ is to implement a subgradient iteration
(a gradient iteration is impossible here because $D(\boldsymbol{\lambda}^R)$
is non-differentiable w.r.t. $[\boldsymbol{\lambda}^R]_m$). Let $\partial
D(\boldsymbol{\lambda}^R)$ denote a subgradient vector of \eqref{E:dual} whose
$m$th entry is $[\partial
D(\boldsymbol{\lambda}^R)]_m:=[\mathbf{\check{r}}]_m-\sum_{\forall\mathbf{J}}\sum_{\forall
k}[\mathbf{R}^*(\mathbf{J},$ $\boldsymbol{\lambda}^R)]_{m,k}$
$[\mathbf{W}^*(\mathbf{J},\boldsymbol{\lambda}^R)]_{m,k}\Pr\{\mathbf{J}\}$; let also $i$ denote an iteration index, and $\beta^{(i)}$ a decreasing small stepsize such that $\sum_{i=1}^{\infty}\beta^{(i)}=\infty$ and $\sum_{i=1}^{\infty}\left(\beta^{(i)}\right)^2<\infty$. With these choices, the iterations
\begin{equation}\label{E:original_off-line-iter}
{\boldsymbol{\lambda}^{R}}^{(i)}={\boldsymbol{\lambda}^{R}}^{(i-1)}+\beta^{(i)}\partial
D({\boldsymbol{\lambda}^R}^{(i-1)})
\end{equation}
converge to $\boldsymbol{\lambda}^{R*}$ as $i\rightarrow\infty$ (cf. \cite[Sec. 6.3.1]{kn:Bertsekas99}).
A major challenge in obtaining $\boldsymbol{\lambda}^{R*}$ using \eqref{E:original_off-line-iter}
is that $[\partial
D(\boldsymbol{\lambda}^R)]_m$ is discontinuous because $\mathbf{W}^*(\mathbf{J},\boldsymbol{\lambda}^R)$
is not continuous for every $\boldsymbol{\lambda}^R$ that gives rise to a tie. This problem is critical, because in most cases $\boldsymbol{\lambda}^{R*}$ is one of the points where $[\partial
D(\boldsymbol{\lambda}^R)]_m$ is discontinuous. Note that discontinuity of the primal solution at $\boldsymbol{\lambda}^{R*}$ implies that obtaining a solution arbitrarily close to the optimal in the dual domain, does not guarantee obtaining a solution arbitrarily close to the optimal in the primal domain. Specifically, after running a sufficiently high but finite number of iterations $I$, we can guarantee that ${\boldsymbol{\lambda}^{R}}^{(I)}$ is a very good approximation for $\boldsymbol{\lambda}^{R*}$, but we cannot guarantee that $\mathbf{W}^*(\mathbf{J},{\boldsymbol{\lambda}^{R}}^{(I)})$ is a good approximation of $\mathbf{W}^*(\mathbf{J},\boldsymbol{\lambda}^{R*})$. In fact, it can be shown that such schedulings are significantly different for a subset of channel realizations $\mathbf{J}$, and that the scheduling $\mathbf{W}^*(\mathbf{J},{\boldsymbol{\lambda}^{R}}^{(I)})$ is not a feasible solution of \eqref{E:optRA} since it violates the average rate constraints.

\begin{figure}
\def\epsfsize#1#2{0.9#1}
\centerline{\epsffile{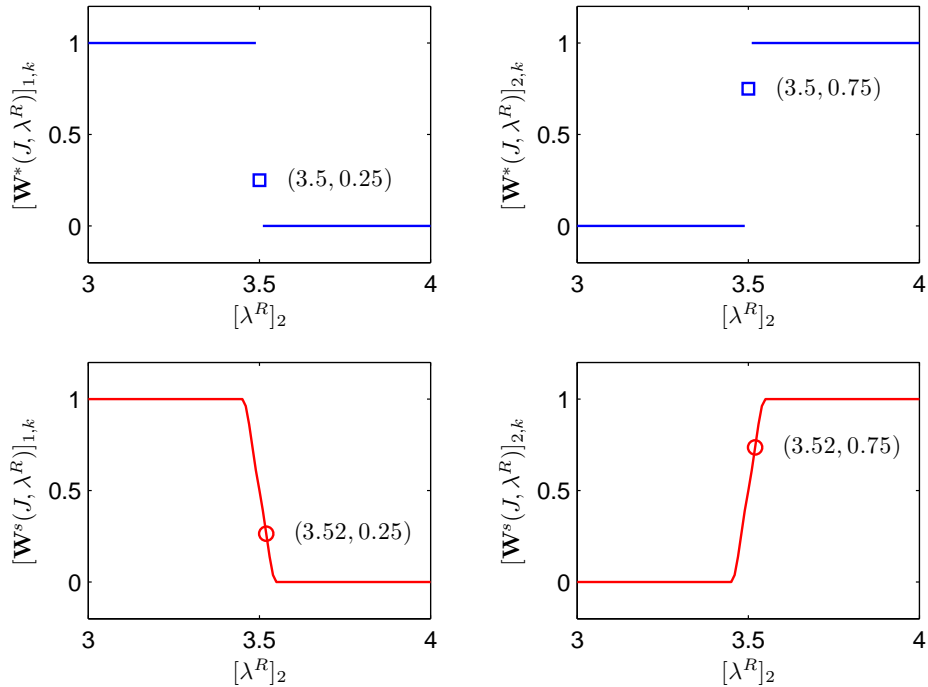}} \caption{Optimal (top) and smooth (bottom) channel allocation for the $k$th channel as $[\boldsymbol{\lambda}^R]_2$ varies. The simulated set-up is: $M=2$, $\varepsilon=0.01$, $[\boldsymbol{\lambda}^R]_1=\lambda_0$ is kept constant, and $[\mathbf{C}_W(\mathbf{J},\boldsymbol{\lambda}^R)]_{1,k}=[\mathbf{C}_W(\mathbf{J},\boldsymbol{\lambda}^R)]_{2,k}$ when $[\boldsymbol{\lambda}^R]_1=\lambda_0$ and $[\boldsymbol{\lambda}^R]_2=3.5$.} \label{F: Ties_Smooth_Hard}
\end{figure}

Our approach to solve this problem is to reinstate Lipschitz continuity by smoothing the scheduling
function. Smoothing ensures continuity or differentiability and has been
successfully applied to different optimization problems;
see e.g., \cite{kn:zenios_smooth} and \cite{Smooth_Nesterov}. Since scheduling discontinuities appear in the transition from
a tie to a single-winner (check \eqref{E:simpl_WW}, \eqref{E:optW_tie} and the left and right upper plots of Figure \ref{F: Ties_Smooth_Hard}), the idea is to relax the condition for
scheduling in the $k$th channel only when $m\in\mathcal{M}(\mathbf{J},k)$.
This is possible through the set $\mathcal{M}^s(\mathbf{J},k):=\{m:
([\mathbf{C}_W(\mathbf{J},\boldsymbol{\lambda}^R)]_{m,k}$
$-[\mathbf{c}_W^*(\mathbf{J},\boldsymbol{\lambda}^R)]_k<\varepsilon)~
\wedge
~([\mathbf{c}_W^*(\mathbf{J},\boldsymbol{\lambda}^R)]_k<0)\}$,
where $\varepsilon$ is a small positive number. Based on
$\mathcal{M}^s(\mathbf{J},k)$, consider the following suboptimal
but smooth scheduling matrix
\begin{gather}\label{E:WSmoth}
\hspace{-3.0cm} [\mathbf{W}^s(\mathbf{J},\boldsymbol{\lambda}^R)]_{m,k}:=\mathbbm{1}_{\{ m\in\mathcal{M}^s(\mathbf{J},k)\}} \frac{\left(1-\frac{[\mathbf{C}_W(\mathbf{J},\boldsymbol{\lambda}^R)]_{m,k}-[\mathbf{c}_W^*(\mathbf{J},
\boldsymbol{\lambda}^R)]_k}{\varepsilon}\right)^2}
        {\sum_{m\in\mathcal{M}^s(\mathbf{J},k)}\left(1-\frac{[\mathbf{C}_W(\mathbf{J},\boldsymbol{\lambda}^R)]_{m,k}
        -[\mathbf{c}_W^*(\mathbf{J},\boldsymbol{\lambda}^R)]_k}{\varepsilon}\right)^2}.
\end{gather}
Clearly, $[\mathbf{W}^s(\mathbf{J},\boldsymbol{\lambda}^R)]_{m,k}$ schedules
channel $k$ not only to users $m$ whose cost is minimum but also to those whose cost is $\varepsilon$-close to
the minimum. This can be readily appreciated in the left lower and right lower plots of the example illustrated in Figure \ref{F: Ties_Smooth_Hard}. According to the upper left plot, when $[\boldsymbol{\lambda}^R]_{2}\in(3.45,3.5)$ the optimum allocation assigns the channel to user 1, meaning that its cost is the lowest in that interval. However, according to the lower right plot, when $[\boldsymbol{\lambda}^R]_{2}\in(3.45,3.5)$ the smooth allocation assigns a portion of the channel also to user 2. This is because although the cost of user 1 is still smaller, within that interval the difference of costs between the two users is less than $\varepsilon$. Something similar happens when $[\boldsymbol{\lambda}^R]_{2}\in(3.5,3.55)$, but in this case user 2 is the one with the smallest cost.

The scheduling in \eqref{E:WSmoth} exhibits other relevant properties that are summarized in the next Proposition.
\begin{proposition}\label{P:propWSmoth}
{\em The smooth scheduler $\mathbf{W}^{s}(\mathbf{J},\boldsymbol{\lambda}^R)$ satisfies the following:\\
(i) If
$[\mathbf{W}^{s}(\mathbf{J},\boldsymbol{\lambda}^R)]_{m,k}>0$, then
$m\in \mathcal{M}^s(\mathbf{J},k)$ and $[\mathbf{C}_W(\mathbf{J},$ $\boldsymbol{\lambda}^R)]_{m,k}<[\mathbf{c}_W^*(\mathbf{J},$$\boldsymbol{\lambda}^R)]_k+\varepsilon$;\\
(ii) If $|\mathcal{M}^s(\mathbf{J},k)|>0$, $\sum_{m\in
\mathcal{M}^s(\mathbf{J},k)}[\mathbf{W}^{s}(\mathbf{J},\boldsymbol{\lambda}^R)]_{m,k}=1$;\\
(iii) If $|\mathcal{M}(\mathbf{J},k)|=0$, then
$[\mathbf{W}^{s}(\mathbf{J},\boldsymbol{\lambda}^R)]_{m,k}=0$
$\forall m$; and\\
(iv) $[\mathbf{W}^s(\mathbf{J},\boldsymbol{\lambda}^R)]_{m,k}$ is a
continuous function of $\boldsymbol{\lambda}^{R}$.}
\end{proposition}
\begin{IEEEproof}
The construction of the scheduling matrix \eqref{E:WSmoth} can be readily used to verify the claims (i)-(iv).
\end{IEEEproof}
Properties \emph{(i)-(iii)} of $\mathbf{W}^s$ are similar to
those of $\mathbf{W}^*$ stated in Proposition \ref{P:propChAll},
while \emph{(iv)} ensures continuity (check lower plots in Figure \ref{F: Ties_Smooth_Hard}). Besides being continuous, the smooth scheduling also lowers complexity relative to its discontinuous counterpart. In fact, when a tie occurs, finding $\mathbf{W}^{*}(\mathbf{J})$ requires solving a linear program that involves channel realizations other than $\mathbf{J}$ (recall Example \ref{Ex:SystFewStates}), while finding $\mathbf{W}^{s}(\mathbf{J})$ requires only the computation of the closed form in \eqref{E:WSmoth} without having to consider any channel realization other than $\mathbf{J}$.

Based on Proposition \ref{P:propWSmoth}, the following result can be established.
\begin{lemma}\label{L:propDSmoth}
{\em If $D^s(\boldsymbol{\lambda}^R):=
\mathcal{L}(\boldsymbol{\lambda}^R,\mathbf{R}^*(\mathbf{J},\boldsymbol{\lambda}^R)\odot\mathbf{W}^s(\mathbf{J},\boldsymbol{\lambda}^R),$
$\mathbf{W}^s(\mathbf{J},\boldsymbol{\lambda}^R))$ and $[\partial^s
D(\boldsymbol{\lambda}^R)]_m:=[\mathbf{\check{r}}]_m-\sum_{\forall\mathbf{J}}\sum_{\forall
k}$ $[\mathbf{R}^*(\mathbf{J},\boldsymbol{\lambda}^R)]_{m,k}$ $[\mathbf{W}^s(\mathbf{J},\boldsymbol{\lambda}^R)]_{m,k}$
$\Pr\{\mathbf{J}\}$ denote smooth versions of the dual function and its subgradient, then:\\
(i) For all $\boldsymbol{\lambda}^R$, it holds that
$D(\boldsymbol{\lambda}^R)\leq D^s(\boldsymbol{\lambda}^R)<D(\boldsymbol{\lambda}^R)+\varepsilon'$, where $\varepsilon':=K\varepsilon$; and \\
(ii) $[\partial^s D(\boldsymbol{\lambda}^R)]_m$ is a Lipschitz
continuous and decreasing function of $\boldsymbol{\lambda}^R$.}
\end{lemma}
\begin{IEEEproof}
Appendix C.
\end{IEEEproof}

Lemma \ref{L:propDSmoth} guarantees that $\partial
D^s(\boldsymbol{\lambda}^R)$ is a Lipschitz continuous
$\varepsilon'$-subgradient of $D(\boldsymbol{\lambda}^R)$ \cite[pp.
625]{kn:Bertsekas99} and will play a critical role in the convergence results presented later in Propositions \ref{P:converg} and \ref{P:online-locking}. At this point, we are ready to prove the following result.
\begin{proposition}\label{P:converg}
{\em If $\beta$ is a small constant stepsize, there exist ${\boldsymbol{\lambda}^{R}}^{(0)}$ so that:\\
(i) the iteration
\begin{equation}\label{E:off-line-iter}
{\boldsymbol{\lambda}^{R}}^{(i)}={\boldsymbol{\lambda}^{R}}^{(i-1)}+\beta\partial^s
D({\boldsymbol{\lambda}^R}^{(i-1)})
\end{equation}
converges, i.e.,
${\boldsymbol{\lambda}^{R}}^{(i)}\rightarrow\boldsymbol{\lambda}^{Rs}$; and \\
(ii) at the limit point it holds that: $D(\boldsymbol{\lambda}^{R*})\leq D^s(\boldsymbol{\lambda}^{Rs})<D(\boldsymbol{\lambda}^{R*})+\varepsilon'$.}
\end{proposition}
\begin{IEEEproof}
To prove part \emph{(i)}, it suffices to show that \eqref{E:off-line-iter} is a nonlinear contraction mapping, which basically requires: (a) existence of $\boldsymbol{\lambda}^{Rs}$ such that $\partial^s D({\boldsymbol{\lambda}^R})=\mathbf{0}$ (this is trivial because the entries of the smooth subgradient are continuous); and (b) the Jacobian of $\partial^s D({\boldsymbol{\lambda}^R})$ to be negative definite with bounded eigenvalues. These two properties of the Jacobian are proved in Appendix D. The proof of part \emph{(ii)} is simpler and relies on Lemma \ref{L:propDSmoth}-(i) and on the fact that there is zero duality gap; see Appendix E for details.
\end{IEEEproof}

Proposition \ref{P:converg} is of paramount importance. First, it guarantees that if $\mathbf{R}^{*}(\mathbf{J},\boldsymbol{\lambda}^R)$ and $\mathbf{W}^{s}(\mathbf{J},\boldsymbol{\lambda}^R)$ are implemented with $\boldsymbol{\lambda}^{R}=\boldsymbol{\lambda}^{Rs}$, then the average rate constraints are satisfied with equality (recall that $\partial^s D({\boldsymbol{\lambda}^R})=\mathbf{0}$ only if this is the case). Second, it provides a systematic algorithm to compute $\boldsymbol{\lambda}^{Rs}$. Third and foremost, it guarantees that the overall weighted average power penalty paid for implementing the smooth policy $\mathbf{R}^{*}(\mathbf{J},\boldsymbol{\lambda}^{Rs})$ and $\mathbf{W}^{s}(\mathbf{J},\boldsymbol{\lambda}^{Rs})$ instead of the optimum policy $\mathbf{R}^*(\mathbf{J},\boldsymbol{\lambda}^{R*})$ and $\mathbf{W}^*(\mathbf{J},\boldsymbol{\lambda}^{R*})$ is less than\footnote{In practice, the gap w.r.t. $D(\boldsymbol{\lambda}^{R*})$
is much smaller than $\varepsilon'$. This is because
$\mathbf{W}^{s}(\mathbf{J},\boldsymbol{\lambda}^R)\neq\mathbf{W}^{*}(\mathbf{J},
\boldsymbol{\lambda}^R)$ only if $|\mathcal{M}^s(\mathbf{J},k)|>1$, which is a rare
event; hence, on average, the bound in Lemma
\ref{L:propDSmoth}-\emph{(i)} is very loose; see also Appendix C.} $\varepsilon'$. The latter assertion is true because according to the definitions of $D(\boldsymbol{\lambda}^{R})$ in \eqref{E:dual} and $D^s(\boldsymbol{\lambda}^{R})$ in Lemma \ref{L:propDSmoth}, the values of the dual functions coincide with those of the Lagrangian in \eqref{E:Lagrangian} when the optimum and the smooth policies are implemented, respectively. Since when $D(\boldsymbol{\lambda}^{R*})$ and $D^s(\boldsymbol{\lambda}^{Rs})$ are evaluated via \eqref{E:Lagrangian} all the constraints are satisfied with equality, the only remaining term in the Lagrangians is the overall weighted average transmitted power. Therefore, the bounds on the dual values in Proposition \ref{P:converg}-(ii), directly translate to bounds on the overall weighted average power consumption.

An algorithm based on Proposition \ref{P:converg} to find $\boldsymbol{\lambda}^{Rs}$ is described next:
\vspace{.2cm}\hrule height0.1pt depth0.3pt \vspace{.2cm}
\noindent\textbf{Algorithm 1}: \emph{Calculation of the Lagrange multipliers}
\vspace{.05cm} \hrule height0.1pt depth0.3pt \vspace{0.3cm}
\noindent \textbf{({S1.0})} \emph{Initialization}: set vectors $\boldsymbol{\delta}_1$,
$\boldsymbol{\delta}_2$ to small positive values;
${\boldsymbol{\lambda}^{R(0)}}=\boldsymbol{\delta}_1$, and the
iteration index $i=1$.\\
\noindent \textbf{({S1.1})} \emph{Resource allocation update}: per Q-CSI realization $\mathbf{J}$,
use ${\boldsymbol{\lambda}^{R}}^{(i-1)}$ to obtain $\mathbf{R(J)}^{(i)}$
and $\mathbf{P(J)}^{(i)}$ based on (\ref{E:rateopt_nom}) and
$\Upsilon_{\mathcal{R}([\mathbf{J}]_{m,k})}$; and
$\mathbf{W}^s(\mathbf{J})^{(i)}$ using (\ref{E:WSmoth}). \\
\noindent \textbf{({S1.2})} \emph{Dual update}: use (S1.1) to find $\partial^s
D({\boldsymbol{\lambda}^R}^{(i-1)})$. Stop if $|\partial^s
D({\boldsymbol{\lambda}^R}^{(i-1)})|< \boldsymbol{\delta}_2$; update
${\boldsymbol{\lambda}^{R}}^{(i)}$ as in (\ref{E:off-line-iter}), and set $i=i+1$; otherwise, go to (S1.1).
\vspace{.1cm}\hrule height0.1pt depth0.3pt \vspace{.5cm}

Due to the average formulation in (\ref{E:optRA}),
Algorithm 1 entails computing the average rate and power per user which require
the knowledge of the joint channel distribution. Specifically, $\Pr\{\mathbf{J}\}$ needs to be known $\forall \mathbf{J}$. It must be run during an initialization (off-line) phase before the communication starts and it only needs to be re-run if either the channel statistics or the users' QoS requirements change. Once $\boldsymbol{\lambda}^{R}$
is known, the ($\varepsilon'$-) optimum allocation per $\mathbf{J}$ is
found online using $\mathbf{R}^{*}(\mathbf{J},\boldsymbol{\lambda}^{Rs})$, $\Upsilon_{\mathcal{R}([\mathbf{J}]_{m,k})}$, and $\mathbf{W}^{s}(\mathbf{J},\boldsymbol{\lambda}^{Rs})$. Since expressions for those are available in closed form [cf. \eqref{E:rateopt_nom} and \eqref{E:WSmoth}], the computational burden associated to the online phase is negligible.

\subsection{Stochastic Estimation of the Lagrange Multipliers}\label{S:stoch}

As mentioned before, $\boldsymbol{\lambda}^{Rs}$ is obtained using Algorithm 1 off-line, and requires knowledge of the channel distribution. However, this computation cannot be always efficiently carried out or may even be infeasible. This is the case when: (a) the number of users, channel statistics, and QoS requirements change so frequently that $\boldsymbol{\lambda}^{R*}$ has to be continuously re-computed; (b) in limited-complexity systems that cannot afford the off-line burden; or (c) when the joint channel distribution is unknown. For those situations, stochastic approximation algorithms \cite{kushner_stoch_approx_book} arise as an alternative solution to estimate $\boldsymbol{\lambda}^{Rs}$ \cite{kn:wang_giann_marq_PIEEE07}. Let $n$ index the
current block (whose duration corresponds to the channel coherence interval $T_{ch}$), and let $\mathbf{J}[n]$ denote the fading state
during block $n$. Our proposal amounts to replace the ensemble average subgradient $[\partial^s
D(\boldsymbol{\lambda}^R)]_m=[\mathbf{\check{r}}]_m-\sum_{\forall
\mathbf{J}}\sum_{\forall
k}[\mathbf{R}(\mathbf{J},\boldsymbol{\lambda}^R))]_{m,k}[\mathbf{W}^s(\mathbf{J},\boldsymbol{\lambda}^R))]_{m,k}\Pr\{\mathbf{J}\}$ with its stochastic version $[\partial^s
D(\boldsymbol{\lambda}^R,n)]_m:=[\mathbf{\check{r}}]_m-\sum_{\forall
k}[\mathbf{R}(\mathbf{J}[n],\boldsymbol{\lambda}^R))]_{m,k}[\mathbf{W}^s(\mathbf{J}[n],\boldsymbol{\lambda}^R))]_{m,k}$. Using this definition\footnote{Stochastic implementations of $\partial^s
D(\boldsymbol{\lambda}^R,n)$ different from the one proposed here are also possible. For example, convergence to the optimum value using arguments similar to those in Proposition \ref{P:online-locking} can be also proved for stochastic versions based on finite time window averaging or sample averaging.}, the original iterations over $\boldsymbol{\lambda}^{R}$ in (\ref{E:off-line-iter}) can be replaced by their estimates
\begin{equation}\label{E:online-iter}
{\boldsymbol{\hat{\lambda}}^{R}}[n+1]={\boldsymbol{\hat{\lambda}}^{R}}[n]+\beta\partial^s
D({\boldsymbol{\hat{\lambda}}^R}[n],n)
\end{equation}
where $\beta$ is again a \emph{constant} stepsize. Capitalizing on the Lipschitz continuity of $\partial^s D(\boldsymbol{\lambda}^R,n)$, it can be shown that for sufficiently small $\beta$: (i) the trajectories of the iterations in (\ref{E:off-line-iter}) and (\ref{E:online-iter}) are locked; and (ii) the stochastic iterates in (\ref{E:online-iter}) converge to a neighborhood of $\boldsymbol{\lambda}^{Rs}$. Specifically, we have:
\begin{proposition}\label{P:online-locking}
{\em With initial conditions similar to (\ref{E:off-line-iter}) and (\ref{E:online-iter}) and given $T>0$, there exist $b_T>0$ and $\beta_T>0$ so that almost surely
\begin{equation}\label{E:locking}
\max_{1\leq n \leq T/\beta }\|{\boldsymbol{\lambda}^{Rs}}^{(n)}-\boldsymbol{\hat{\lambda}}^{Rs}[n]\|\leq c_T(\beta)b_T \hspace{1cm}
\end{equation}
where $0\leq\beta\leq \beta_T$ and $c_T(\beta)\rightarrow 0$ as $\beta\rightarrow 0$.}
\end{proposition}
\begin{IEEEproof}
The result in \eqref{E:locking} can be shown by adopting the averaging approach in
\cite[Chapter 9]{SoloBook}. Following the
averaging method for approximating the difference equation trajectory, the updates in
\eqref{E:online-iter} and those in
\eqref{E:off-line-iter} can be seen as a pair
of {\em primary} and {averaged} systems. Under general conditions,
it is possible to show the trajectory locking of these two systems
via \cite[Theorem 9.1]{SoloBook}. The full proof of the proposition is omitted due to space limitations, but the main idea hinges on the Lipschitz continuity of $\partial^s D(\boldsymbol{\lambda}^R,n)$ to prove that the most challenging conditions required in \cite[Theorem 9.1]{SoloBook} hold. Interestingly, as $n\rightarrow\infty$ a similar approach can be used to show convergence in probability of \eqref{E:online-iter} to \eqref{E:off-line-iter}, \cite[Theorem 9.5]{SoloBook}.
\end{IEEEproof}

Proposition \ref{P:online-locking} not only states that the trajectories of the online iterations remain locked to those of the original ensemble (off-line) iterations, but also that the gap between those shrinks as the stepsize (that is at our disposal) vanishes. The result holds for a constant (non-zero) $\beta$, which allows the iterations in \eqref{E:online-iter} to cope with channel non-stationarities and track changes in the system set-up (e.g., users entering or leaving the system). This type of convergence is different from that exhibited by other relevant stochastic resource allocation schemes \cite{Stolyar05QS}, \cite{kn:wang_giann_marq_PIEEE07}.

From an implementation perspective, it must be emphasized that iterations in (\ref{E:online-iter}) can be implemented online without knowing the channel distribution. This eliminates the need for implementing Algorithm 1 during an off-line phase, and greatly reduces the overall complexity. However, they moderately increase the complexity during the online (communication) phase. To clarify these assertions, a description of the system operation when the channel-adaptive schemes are implemented based on $\boldsymbol{\lambda}^{Rs}$ (non-stochastic implementation) and when those schemes are implemented based on $\boldsymbol{\hat{\lambda}}^R[n]$ (stochastic implementation) is presented next.
\begin{itemize}
\item
Systems implementing non-stochastic adaptive schemes operate in two phases. During an off-line (initialization) phase Algorithm 1 is executed and the returned value of $\boldsymbol{\lambda}^{Rs}$ is distributed to the transceivers. During the online phase, the value of $\mathbf{J}$ is updated every coherence interval, and the powers, rates and scheduling are adapted with $\boldsymbol{\lambda}^{R}=\boldsymbol{\lambda}^{Rs}$ and $\mathbf{J}=\mathbf{J}[n]$.
\item Systems implementing stochastic adaptive schemes operate purely online. During the online phase \emph{two} tasks are implemented per coherence interval. First, the powers, rates and scheduling are adapted with $\boldsymbol{\lambda}^{R}={\boldsymbol{\hat{\lambda}}^{R}}[n]$ and $\mathbf{J}=\mathbf{J}[n]$. Second, the multipliers estimates for the next block ${\boldsymbol{\hat{\lambda}}^{R}}[n+1]$ are updated according to (\ref{E:online-iter}).
\end{itemize}

The stochastic schemes also entails change in the place where computations are implemented. For the non-stochastic case, Algorithm 1 will likely be implemented at the access point and the value of $\boldsymbol{\lambda}^{Rs}$ will be transmitted once wherever needed. However, for the stochastic case, $\boldsymbol{\lambda}^{Rs}[n]$ is updated every coherence interval, and therefore instantaneous broadcasting of the analog value of $\boldsymbol{\lambda}^{Rs}[n]$ is not feasible. This implies that during the system operation, iterations in \eqref{E:online-iter} will have to be implemented at different locations. This way, a transmitter that wishes to implement its optimal rate loading in \eqref{E:rateopt_nom} will need to know its own entry of $\boldsymbol{\hat{\lambda}}^R[n]$, while an access point that wants to find the optimum scheduling in \eqref{E:WSmoth} will need to know the value of the entire $\boldsymbol{\hat{\lambda}}^R[n]$. As Proposition \ref{P:online-locking} states, to ensure consistency all the transceivers will have to use identical initialization.

\section{Overhead Issues}\label{S:Overhead_Red}

Previous sections focused on the formulation of the channel-adaptive schemes as well as on developing systematic ways to obtain the variables involved in these optimal schemes. The overhead involved in such schemes is the main goal of this section which relates to practical implementation issues. Specifically, we try to answer questions as: What is the number of different optimum resource allocations? What is the amount of feedback required to implement the developed schemes? How do the functions involved in the optimal schemes look for practical modulations? This overview not only will allow for more efficient implementations of the novel adaptive schemes but also will provide insight to better understand channel-adaptive resource allocation and finite-rate feedback.

\subsection{Exploiting the structure of the optimum solution}\label{S:struct-cardin}

Two properties of the optimal resource allocation are useful to reduce the computational overhead. Specifically, we observe that:
\begin{enumerate}[{\em P1)}]
\item
Given $\boldsymbol{\lambda}^{R*}$, the optimum rate matrix $\mathbf{R}^*$ in (\ref{E:rateopt_nom}) satisfies the following: (i) for a given user $m$ it does not depend on the
other users $m'\neq m$; and (ii) the
optimum rate allocation for channel $k$ can be carried out
separately from the allocation of the remaining $k'\neq k$
channels. Since the power-rate function depends on the specific region $\mathcal{R}([\mathbf{J}]_{m,k})$, the previous properties imply that the
optimal rate (and thus power) allocation for user $m$ on channel $k$
can be obtained separately from the rate allocation in the
remaining regions $\mathcal{R}([\mathbf{J}]_{m,k}')\neq\mathcal{R}([\mathbf{J}]_{m,k})$. In other words, the rate allocation can be written as $[\mathbf{{R}^{*}}(\mathbf{J})]_{m,k}=[\mathbf{{R}^{*}}([\mathbf{J}]_{m,k})]_{m,k}$.
%
\item \label{L:pp3}
Given $\boldsymbol{\lambda}^{R*}$, the previous observations can be used to obtain the cost indicator function as $[\mathbf{C}_W(\mathbf{J})
]_{m,k}=[\mathbf{C}_W([\mathbf{J}]_{m,k})]_{m,k}$ $\forall
(\mathbf{J},m,k)$. Since the user scheduling for channel $k$, that is $[\mathbf{W}^s(\mathbf{J})]_{m,k}$ $\forall m$, is found based on $[\mathbf{C}_W([\mathbf{J}]_{m,k})]_{m,k}$ $\forall m$, information about channels $k'\neq k$ is not needed [c.f. \eqref{E:WSmoth}]. Therefore, the user-scheduling allocation can be written as $[\mathbf{W}^s(\mathbf{J})]_{m,k}=[\mathbf{W}^s([\mathbf{J}]_{k})]_{m,k}$.
\end{enumerate}
Properties \emph{P1)} and \emph{P2)} point out that for a given channel realization $\mathbf{J}$, vector $\boldsymbol{\lambda}^{R*}$ encapsulates most of the information the $(m,k)$ user-channel pair needs from: channel realizations different than $\mathbf{J}$, channels different than $k$, and users different than $m$.

To appreciate the implications of \emph{P1)} and \emph{P2)}, in the following we will consider that each individual channel domain is divided into $L$ quantization regions. Without
loss of optimality the quantization regions can be represented by a set of
thresholds $\{q_{m,k,l}\}_{l=1}^{L+1}$ \cite{kn:amfdgg08}. Hence, if $g_{m,k}\in[q_{m,k,l},q_{m,k,l+1})$, then $[\mathbf{J}]_{m,k}=l$; see e.g.~\cite{kn:amfdgg08}. (Note that since $g_{m,k}\in \mathbb{R}_+$,
$q_{m,k,1}=0$ and $q_{m,k,L+1}=\infty$ $\forall(m,k)$.)

An immediate implication of \emph{P1)} and \emph{P2)} is that the average over $\mathbf{J}$ can be decomposed into sub-averages across channels. Specifically, with $\mathcal{J}_k$ denoting the set of possible values
$[\mathbf{J}]_k$ takes, each individual average rate can be rewritten as
$${\displaystyle \sum_{\forall \mathbf{J} \in
\mathcal{J}}}\left(\sum_{k=1}^{K} [{\mathbf{R}^*([\mathbf{J}])}]_{m,k}[\mathbf{W}^*(\mathbf{J})]_{m,k}\right)\Pr\{\mathbf{J}\}={\displaystyle
\sum_{k=1}^{K}}\left({\sum_{\forall \mathbf{j} \in \mathcal{J}_k}}
[{\mathbf{R^*}(\mathbf{j})}]_{m,k}[\mathbf{W}^*(\mathbf{j})]_{m,k}\Pr\{[\mathbf{J}]_k=\mathbf{j}\}\right).$$
While the left hand side requires $K|\mathcal{J}|=K L^{KM}$ summations, the right hand side only requires $K|\mathcal{J}_k|=K L^{M}$.

 Another possibility to reduce complexity is to cluster different channel realizations that give rise to the same optimal resource allocation. For example, consider a channel realization $\mathbf{J}_1$ for which user $m'$ is found to be the winner for the $k$th channel, and a different channel realization $\mathbf{J}_2$ so that $[\mathbf{J}_1]_{m',k}=[\mathbf{J}_2]_{m',k}$ and $[\mathbf{C}_W(\mathbf{J}_2)]_{m,k}>[\mathbf{C}_W(\mathbf{J}_1)]_{m,k}$ $\forall m\neq m'$. It is clear that user $m'$ will be again the winner and the resource allocation over the $k$th channel for both $\mathbf{J}_1$ and $\mathbf{J}_2$ will be the same. This can be formalized as follows.
\begin{proposition}\label{P:cq4}
\emph{Assume that $[\mathbf{R}^*([\mathbf{J}]_{m,k}+1)]_{m,k}\geq[\mathbf{R}^*([\mathbf{J}]_{m,k})]_{m,k}$ (i.e., the better the channel the higher the allocated rate), and define $\mathcal{J}_k^{m,l}:=\{\mathbf{j}\in\mathcal{J}_k:
~[\mathbf{W}^*(\mathbf{j})]_{m,k}=1~\wedge~[\mathbf{j}]_m=l \}$. It then holds that:}
\begin{enumerate}[{\it (i)}]
\item {\it $~~$ If $\mathbf{j}\in \mathcal{J}_k^{m,l}$, then
$\{\mathbf{j}'\in
\mathcal{J}_k:~[\mathbf{j}']_{m'}=[\mathbf{j}]_{m'}~\forall
m'\neq m~\wedge~[\mathbf{j}']_m\geq[\mathbf{j}]_m\}\subseteq
\mathcal{J}_k^{m,l}$}

\item {\it $~~$ If $\mathbf{j}\in \mathcal{J}_k^{m,l}$, then
$\{\mathbf{j}'\in
\mathcal{J}_k:~[\mathbf{j}']_{m'}\leq[\mathbf{j}]_{m'}~\forall
m'\neq m~\wedge~[\mathbf{j}']_m=[\mathbf{j}]_m\}\subseteq
\mathcal{J}_k^{m,l}$}

\item {\it $~~$ If $\mathbf{j}\notin \mathcal{J}_k^{m,l}$, then
$\{\mathbf{j}'\in
\mathcal{J}_k:~[\mathbf{j}']_{m'}\geq[\mathbf{j}]_{m'}~\forall
m'\neq m~\wedge~[\mathbf{j}']_m=[\mathbf{j}]_m\}\nsubseteq
\mathcal{J}_k^{m,l}$}
\end{enumerate}
\end{proposition}
\begin{IEEEproof}
Appendix F.
\end{IEEEproof}
Under the reasonable assumption that $[\mathbf{R}^*([\mathbf{J}]_{m,k}+1)]_{m,k}\geq[\mathbf{R}^*([\mathbf{J}]_{m,k})]_{m,k}$ (which is true for the examples of $\Upsilon$ in this paper), the properties in Proposition \ref{P:cq4} allow one to group the channel realizations $\mathbf{J}$ in clusters, which yield the same
optimum resource allocation. Clustering can be exploited to reduce the calculations
required to determine the optimum resource allocation (Algorithm 1) as well as to reduce the finite-rate feedback overhead as discussed next.

\subsection{Finite-Rate Feedback}\label{S:FRF}

As it was mentioned in Section \ref{S:Introduction}, for non-reciprocal channels the Q-CSI
can be naturally obtained at the transmitters through finite-rate feedback from the
receiver. Since $\mathcal{J}$ has finite cardinality, clearly a finite
number of bits $B:=\lceil\log_2(|\mathcal{J}|)\rceil$ suffices to index the current realization $\mathbf{J}$. To
ensure that the Q-CSIT coincides with
the Q-CSIR we will assume that:\\
{\bf ({as2})} \emph{the feedback channel
is error-free, incurs negligible delay, and the channels remain
invariant over at least two consecutive symbols}.\\
Note that this is a pragmatic assumption for Q-CSI since each channel can
vary from one symbol to the next so long as the quantization region
it falls into remains invariant. In addition, error-free feedback is typically
guaranteed with sufficiently strong error control codes especially
since rate in the reverse link is low.

Although in principle the resource allocation varies as a function of $\mathbf{J}$, it is important to note that from an operational perspective the main objective is not feeding back the current
$\mathbf{J}$ to the transmitters, but identifying the optimal resource allocation the transmitters have to implement. These tasks are not equivalent because as it was stated in Proposition \ref{P:cq4}, different channel realizations can be mapped to the same resource allocation. In other words, although a receiver actually realizes that the quantized value of the channel has changed from $\mathbf{J}_1$ to $\mathbf{J}_2$, if the resource allocation is the same in both cases, for the transmitters there is no difference between $\mathbf{J}_1$ and $\mathbf{J}_2$ and they do not need feedback from the receiver notifying them that the channel has changed. This is a meaningful difference because, as it was hinted by \emph{P1)} and \emph{P2)}, the cardinality of the optimal resource allocation is much smaller than the cardinality of the Q-CSI matrix. Therefore, in order to find the minimum amount of feedback the transmitters require, the cardinality of the optimum resource allocation, $[\mathbf{R}^*(\mathbf{J})]_{m,k}=[\mathbf{R}^*([\mathbf{J}]_{m,k})]_{m,k}$ and $[\mathbf{W}^s(\mathbf{J})]_k=[\mathbf{W}^s([\mathbf{J}]_{k})]_k$, has to be carefully examined.

Regarding the rate (power) allocation, it easy to see that $|\{[\mathbf{R}^*([\mathbf{J}]_{m,k})]_{m,k}\}_{\forall
\mathbf{J}}|=L$. The cardinality of the set of different user schedulings depends on whether the winner is unique or not. The cardinality when the winner is unique is also easy to decipher: either $|\{[\mathbf{W}^s([\mathbf{J}]_{k})]_{k}\}_{\forall
\mathbf{J}}|=M$ if there is always one user active, or, $|\{[\mathbf{W}^s([\mathbf{J}]_{k})]_{k}\}_{\forall
\mathbf{J}}|=M+1$ if the additional case of ``no-user-transmitting'' is considered (i.e., the possibility that $|\mathcal{M}(\mathbf{J},k)|=0$).
For those channel realizations for which the winner is non-unique the analysis is more complicated. Consider again the system described in Example \ref{Ex:SystFewStates} with $K=1$ and $M=4$, and suppose now that we have a channel realization $\mathbf{J'}=[\mathbf{J'}]_1$ so that user 1 achieves the minimum cost $[\mathbf{C}_W(\mathbf{J'})]_{1,1}$, but the cost of user 2 is very close to it, e.g.,  $[\mathbf{C}_W(\mathbf{J'})]_{2,1}=[\mathbf{C}_W(\mathbf{J'})]_{1,1}+\varepsilon/2$. Substituting those costs into \eqref{E:WSmoth}, we have $[\mathbf{W}^{s}(\mathbf{J'})]_{1,1}=4/5$ and $[\mathbf{W}^{s}(\mathbf{J'})]_{1,1}=1/5$. This implies that the set $\{\mathbf{W}^s(\mathbf{J})\}_{\forall
\mathbf{J}}$ not only contains the single-user allocations $\{[1,0,0,0]^T,[0,1,0,0]^T,[0,0,1,0]^T,[0,0,0,1]^T,[0,0,0,0]^T\}$, but also the additional element $[4/5,1/5,0,$ $0]^T$. From a practical perspective, it is worth noticing that the user-sharing policy can be implemented in two different ways. Recalling that $T_{ch}$ denotes the coherence interval a first option is for user 1 to transmit during $T_{ch}(4/5)$ seconds and user 2 during the remaining $T_{ch}/5$ seconds. Alternatively, each time that realization $\mathbf{J}$ occurs, user 1 can transmit with probability 4/5 and user 2 transmits in the remaining cases. Note that if scheduling is implemented following the first option, the number of different user schedulings per channel is indeed higher than $M+1$. However, if the system implements the second option the cardinality of the different user-scheduling policies is $|\{[\mathbf{W}^s([\mathbf{J}]_{k})]_{k}\}_{\forall
\mathbf{J}}|=M+1$, maintaining its original value. Since the second implementation entails lower feedback overhead, in the ensuing analysis it will be assumed that the system implements channel sharing using a probabilistic access scheme.

Based on the previous observations, for the receiver to notify the transmitters of the optimum resource allocation, the following information has to be fed back per channel: the index of the winner user index ($M$ possibilities) together with the index of the rate (and power) allocation for that user ($L$ possibilities), plus an additional codeword corresponding to the event of no-user transmitting. This implies that the total feedback required per channel is $\lceil \log_2(ML+1)\rceil$ bits. Since the resource allocation is not coupled across channels, the total amount of feedback required is $B'=\lceil K \log_2(ML+1) \rceil$ bits. This number is significantly smaller than that required to identify the specific channel realization, $\lceil\log_2(|\mathcal{J}|)\rceil=\lceil K\log_2(L^M)\rceil$ bits. In other words, the receiver does not have to index the quantized version of the channel, but the quantized version of the \emph{channel state} information.

Finally, it is worth remarking that the assessment of overhead so far does not exploit the potential correlation of the fading channel across users (i.e., $[\mathbf{J}^T]_{m}$ and $[\mathbf{J}^T]_{m'}$), channels (i.e., $[\mathbf{J}]_{k}$ and $[\mathbf{J}]_{k'}$), or time (i.e., $\mathbf{J}[n]$ and $\mathbf{J}[n']$). If those were considered, the total amount of feedback could be further reduced. Although exploiting the channel correlation to reduce the feedback overhead is certainly a topic of interest, it goes beyond the scope of this work.

\subsection{A simple channel model}\label{S:AssumptionChannelPDF}

In this section, several assumptions that allow one to obtain explicit
expressions for the probability mass function of the channel are made. Suppose first that:\\
{\bf ({as3})} {\it the fading processes for different users are
uncorrelated, which implies that $\mathbf{J}$ has uncorrelated columns; and} \\
{\bf ({as4})} {\it user channels are allowed to be correlated, and
each is complex Gaussian distributed; that is, if$~\overline{g}_{m,k}$
denotes the average channel gain,
$f_{g_{m,k}}(g_{m,k})=(1/\overline{g}_{m,k}) \exp
(-g_{m,k}/\overline{g}_{m,k})$ is the exponential pdf of
$g_{m,k}$}.\\
Note that ({as3}) is common when the users are scattered along space, while ({as4})
corresponds to a Rayleigh flat fading model.

Using (as3), (as4), and the fact that quantization regions for individual channel gains are represented by the set of thresholds $\{q_{m,k,l}\}_{l=1}^{L+1}$, the probabilities
$\Pr\{[\mathbf{J}]_{m,k}=j_{m,k}\}$ and $\Pr\{[\mathbf{J}]_{k}=\mathbf{j}\}$ can be respectively found as
\begin{eqnarray}\label{E:Prob_Ass34}
\Pr\{[\mathbf{J}]_{m,k}=j_{m,k}\}&=&e^{-\frac{q_{m,k,j_{m,k}}}{\overline{g}_{m,k}}}-e^{-\frac{q_{m,k,j_{m,k}+1}}{\overline{g}_{m,k}}}\\
\Pr\{[\mathbf{J}]_{k}=\mathbf{j}\}&=&\prod_{m=1}^{M}\left(e^{-\frac{q_{m,k,[\mathbf{j}]_m}}{\overline{g}_{m,k}}}-e^{-\frac{q_{m,k,[\mathbf{j}]_m+1}}{\overline{g}_{m,k}}}\right).
\end{eqnarray}

\subsection{Examples of power-rate functions}\label{S:examplesofUpsilon}

Another issue affecting implementation aspects of the developed schemes concerns the scenarios for which the power-rate function $\Upsilon(x)$ satisfies (as1). Using Shannon's capacity formula, expressions for $\Upsilon(x)$ and $\Upsilon^{-1}(x)$ that for every region guarantee a specific outage capacity were given in Example \ref{Fot:footnote_BER}. If instead of that definition, one considers the ergodic capacity of user $m$ over the $k$th channel for its $[\mathbf{J}]_{m,k}$th region, it follows that $r_{m,k}=\int_{g_{m,k}\in\mathcal{R}([\mathbf{J}]_{m,k})}\log_2(1+p_{m,k}g_{m,k})f_{g_{m,k}}(g_{m,k})dg_{m,k}$. Using (as4), $\Upsilon^{-1}(x)$ and implicitly $\Upsilon(x)$ can be written as:
\begin{eqnarray}
\Upsilon_{\mathcal{R}([\mathbf{J}]_{m,k})}^{-1}\left(x\right)&=&\int_{q_{m,k,[\mathbf{J}]_{m,k}}}^{q_{m,k,[\mathbf{J}]_{m,k}+1}}\log_2(1+xg_{m,k})\frac{e^{-g_{m,k}/\bar{g}_{m,k}}}{\bar{g}_{m,k}\Pr\{[\mathbf{J}]_{m,k}\}}dg_{m,k}\label{E:r-p_ErgCap}\\
\Upsilon_{\mathcal{R}([\mathbf{J}]_{m,k})}&=&\left\{\vphantom{\frac{\frac{1^^1}{1^^1}}{\frac{1^^1}{1^^1}}}
x\rightarrow y: x-\Upsilon_{\mathcal{R}([\mathbf{J}]_{m,k})}^{-1}\left(y\right)=0\right\}\label{E:p-r_ErgCap}.
\end{eqnarray}
If convenient, the exponential integral function $E_1(x):=\int_x^\infty\exp(-t)/t dt$ can be used to re-write \eqref{E:r-p_ErgCap} in closed form as:
\begin{eqnarray}
\nonumber\Upsilon_{\mathcal{R}([\mathbf{J}]_{m,k})}^{-1}\left(x\right)&=&\left[\log(1+xq_{m,k,[\mathbf{J}]_{m,k}})
e^{\frac{-q_{m,k,[\mathbf{J}]_{m,k}}}{\bar{g}_{m,k}}}
+E_1\left(\frac{1+xq_{m,k,[\mathbf{J}]_{m,k}}}{x\bar{g}_{m,k}}\right)e^{\frac{1}{x\bar{g}_{m,k}}}\right.\\
\nonumber  &-&\left.\log(1+xq_{m,k,[\mathbf{J}]_{m,k}+1})e^{\frac{-q_{m,k,[\mathbf{J}]_{m,k}}}{\bar{g}_{m,k}}}
-E_1\left(\frac{1+xq_{m,k,[\mathbf{J}]_{m,k}+1}}{x\bar{g}_{m,k}}\right)e^{\frac{1}{x\bar{g}_{m,k}}}\right]\\
&\times& \log_2(e)\left[e^{-\frac{q_{m,k,[\mathbf{J}]_{m,k}}}{\overline{g}_{m,k}}}-e^{-\frac{q_{m,k,[\mathbf{J}]_{m,k}+1}}{\overline{g}_{m,k}}}\right]^{-1}.
\end{eqnarray}
Since $\Upsilon^{-1}(x)$ is monotonically increasing [cf. \eqref{E:r-p_ErgCap}], it readily follows that $\Upsilon(x)$ is also monotonically increasing. The strict convexity of $\Upsilon(x)$ is shown in Appendix G.

Besides the power-rate relationship given by the capacity formula, there are situations where transmissions are implemented using pre-specified coding and modulation schemes. Since in those cases a maximum BER is typically prescribed, it is possible to use the BER requirement in order to relate power and rate over a given region. To be more specific, suppose that:\\
{\bf ({as5})} {\it the symbols are drawn from coded modulations
such that the BER function can be adequately approximated by
$\epsilon(g_{m,k},p_{m,k},r_{m,k})\simeq\kappa_1\exp{\left(-g_{m,k}p_{m,k}\kappa_2/(2^{r_{m,k}}-1) \right)}$},\\
where $\kappa_1$ and $\kappa_2$ are constants that depend on the specific modulation and code implemented (e.g., for the uncoded case we typically have $\kappa_2=1$). In addition to being accurate for many practical modulations
\cite{GoldsmithBook} and \cite{kn:goldsmith_vrvp}, (as5) yields tractable mathematical expressions.

If QoS requirements impose a maximum \emph{instantaneous}
BER $\epsilon_{\max}$ per user, ({as5}) can be used to obtain
$\Upsilon(x)$ in explicit
form as
\begin{equation}\label{E:up_BER_max}
\Upsilon_{\mathcal{R}([\mathbf{J}]_{m,k})}\left(x\right)=\frac{(2^{x}-1)
\ln(\kappa_1/\epsilon_{\max})}{\kappa_2q_{m,k,[\mathbf{J}]_{m,k}}}.
\end{equation}
Note that if a powerful coding scheme giving rise to a coding gain of $\kappa_2=\ln(\kappa_1/\epsilon_{\max})$ is implemented, then \eqref{E:up_BER_max} reduces to the one introduced in Example \ref{Fot:footnote_BER} that was derived from the formula of the outage capacity for $\delta=0$. The adoption of maximum instantaneous BER as a QoS requirement also implies that the first region will always represent an outage region with zero power and rate since the power cost for transmitting even minimal rate is infinite.

If QoS requirements dictate that for every region, channel and user a maximum \emph{average} BER $\overline{\epsilon}$ can be tolerated, then $\Upsilon(x)$ is an implicit
function
\begin{eqnarray}\label{E:up_BER_av}
\nonumber\Upsilon_{\mathcal{R}([\mathbf{J}]_{m,k})}&=&\left\{x\rightarrow y:\overline{\epsilon}=\int_{q_{m,k,[\mathbf{J}]_{m,k}}}^{q_{m,k,[\mathbf{J}]_{m,k}+1}}
\epsilon(g_{m,k},y,x)\frac{e^{-g_{m,k}/\bar{g}_{m,k}}}{\bar{g}_{m,k}\Pr\{[\mathbf{J}]_{m,k}\}}dg_{m,k}\right\}\\
&=&\left\{\vphantom{\frac{\frac{1^^1}{1^^1}}{\frac{1^^1}{1^^1}}}
x\rightarrow y:
~\frac{\overline{\epsilon}}{\kappa_1}=\right. \left.\frac{e^{-\frac{\kappa_2q_{m,k,[\mathbf{J}]_{m,k}-1}}{\overline{g}_{m,k}}\left(1+\frac{y\overline{g}_{m,k}}{2^x-1}\right)}-e^{-\frac{\kappa_2q_{m,k,[\mathbf{J}]_{m,k}}}{\overline{g}_{m,k}}\left(1+\frac{y\overline{g}_{m,k}}{2^x-1}\right)}}
{
\left(e^{-\frac{\kappa_2q_{m,k,[\mathbf{j}]_m-1}}{\overline{g}_{m,k}}}
-e^{-\frac{\kappa_2q_{m,k,[\mathbf{j}]_m}}{\overline{g}_{m,k}}}\right)
\left(1+\frac{y\overline{g}_{m,k}}{2^x-1}\right)}\right\}.
\end{eqnarray}
It can be shown that $\Upsilon(x)$ can be written as an
explicit function of the \emph{optimum} rate, $[\boldsymbol{\mu}]_m$ and $[\boldsymbol{\lambda}^{R}]_m$ as
\begin{equation}\label{E:up_BER_av_exp}
{\Upsilon}_{\mathcal{R}([\mathbf{J}]_{m,k})}\left(x\right)=\frac{(2^{x}-1)
[\boldsymbol{\lambda}^{R}]_m}{2^{x}\ln(2)[\boldsymbol{\mu}]_m}.
\end{equation}
Convexity of (\ref{E:up_BER_max}) and
(\ref{E:up_BER_av}) is established in Appendix G. Clearly, alternative ${\Upsilon}(x)$ functions satisfying (as1) can be
derived for modulations whose BER does not satisfy (as5). For example, any $\epsilon(g_{m,k},p_{m,k},r_{m,k})$ that is increasing w.r.t. $r_{m,k}$ and decreasing w.r.t. $p_{m,k}$ while being jointly convex w.r.t. $p_{m,k}$ and $r_{m,k}$ will give rise to a strictly convex ${\Upsilon}(x)$.

From an implementation perspective, not having
$\Upsilon_{\mathcal{R}([\mathbf{J}]_{m,k})}$ in closed form (thus not having
$\dot{\Upsilon}_{\mathcal{R}([\mathbf{J}]_{m,k})}^{-1}$ in closed form) does not necessarily incur a major penalty in terms of computational complexity. Since those expressions do not change with time, the computational burden can be reduced by characterizing those over the domain of interest only once, and using those characterizations for each iteration.

\section{Numerical Examples}\label{S:Simulations}
To test the algorithms developed, we simulated uncorrelated complex
Gaussian fading channels per user adhering to (as2) and (as3), and quantized each channel
gain $g_{m,k}$ to $L_{m,k}=L=4$ regions using the low-complexity channel quantizer in \cite[Sec. IV.B]{kn:amfdgg08}. The power-rate function considered is $\Upsilon_{\mathcal{R}([\mathbf{J}]_{m,k})}\left(x\right)=((2^{x}-1)/g_{m,k}^{\min}([\mathbf{J}]_{m,k})$, derived from the outage capacity formula in Example \ref{Fot:footnote_BER}. Recall that as discussed in Section \ref{S:examplesofUpsilon}, a properly scaled version of this function is also valid for a maximum instantaneous BER requirement [cf. \eqref{E:up_BER_max}].

\begin{figure}
\def\epsfsize#1#2{0.62#1}
\centerline{\epsffile{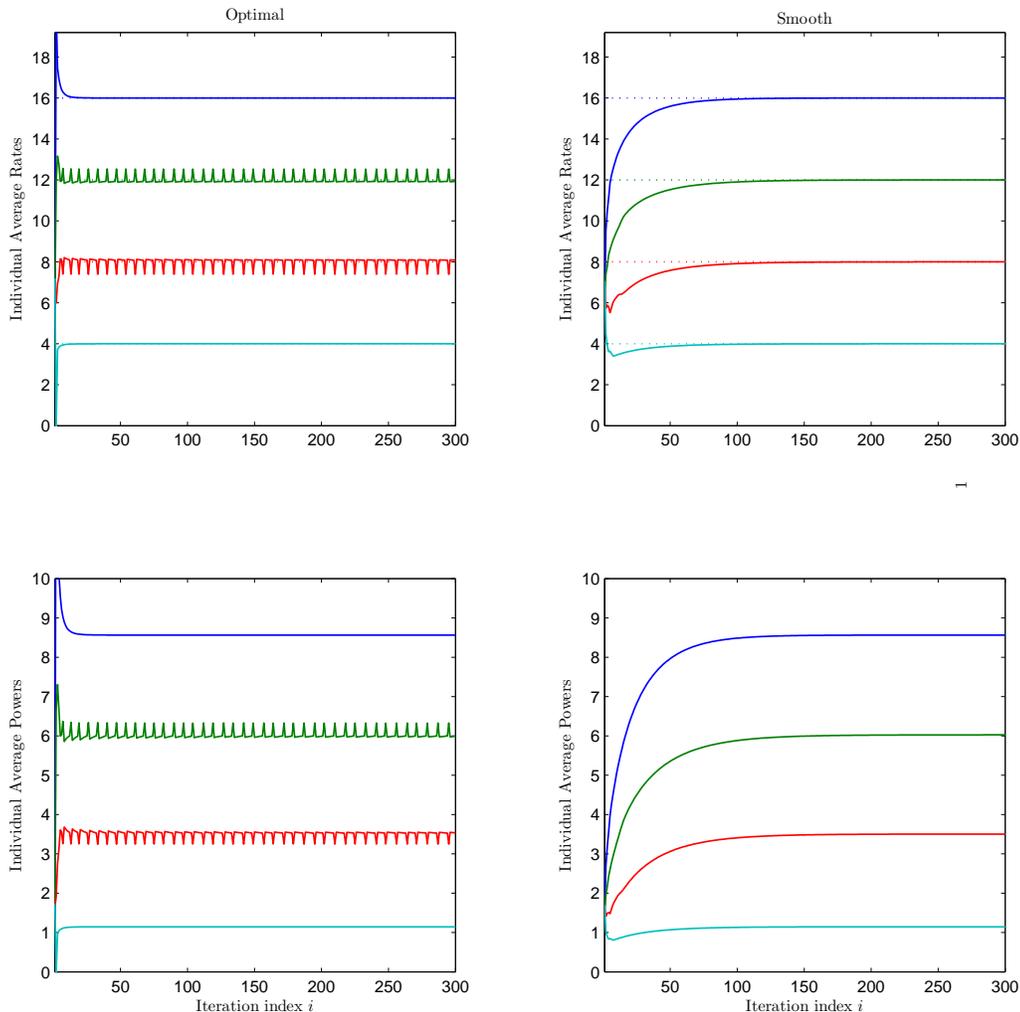}} \caption{Trajectories of average
transmit-rates (top) and transmit powers (bottom) for off-line iterations. The iterations based on the optimal non-smooth policy are shown in the left while the iterations based on the smooth policy are shown in the right.} \label{F:Converg_Offline_Rates_Powers}
\end{figure}

\noindent \textbf{Test Case 1} (Convergence of off-line iterations):
A time-division multiple access (TDMA) system was simulated with $K=16$
uncorrelated channels to serve $M=4$ users with minimum rate requirements
$\mathbf{\check{r}}=[4, 8, 12, 16]$ with an average SNR of $6$dB. Upper plots in Figure \ref{F:Converg_Offline_Rates_Powers}
depict average individual rates versus off-line iterations
for: (i) the subgradient iteration based on the optimal policies in \eqref{E:original_off-line-iter} with $\beta^{(i)}=\kappa i^{0.51}$ (left top); and (ii) the iterations based on the smooth policies in \eqref{E:off-line-iter} with $\varepsilon=0.05$ and $\beta=10^{-2}$ (right top). The trajectories confirm that while the iterations based on the optimal scheduling do not always satisfy the constraints and rate allocation hovers around its
optimum, the smooth policy converges in a finite number of iterations. Behavior of the trajectories of transmit-powers shown in the lower plots of Figure \ref{F:Converg_Offline_Rates_Powers} is similar to that for transmit-rates.

\begin{figure}
\def\epsfsize#1#2{0.62#1}
\centerline{\epsffile{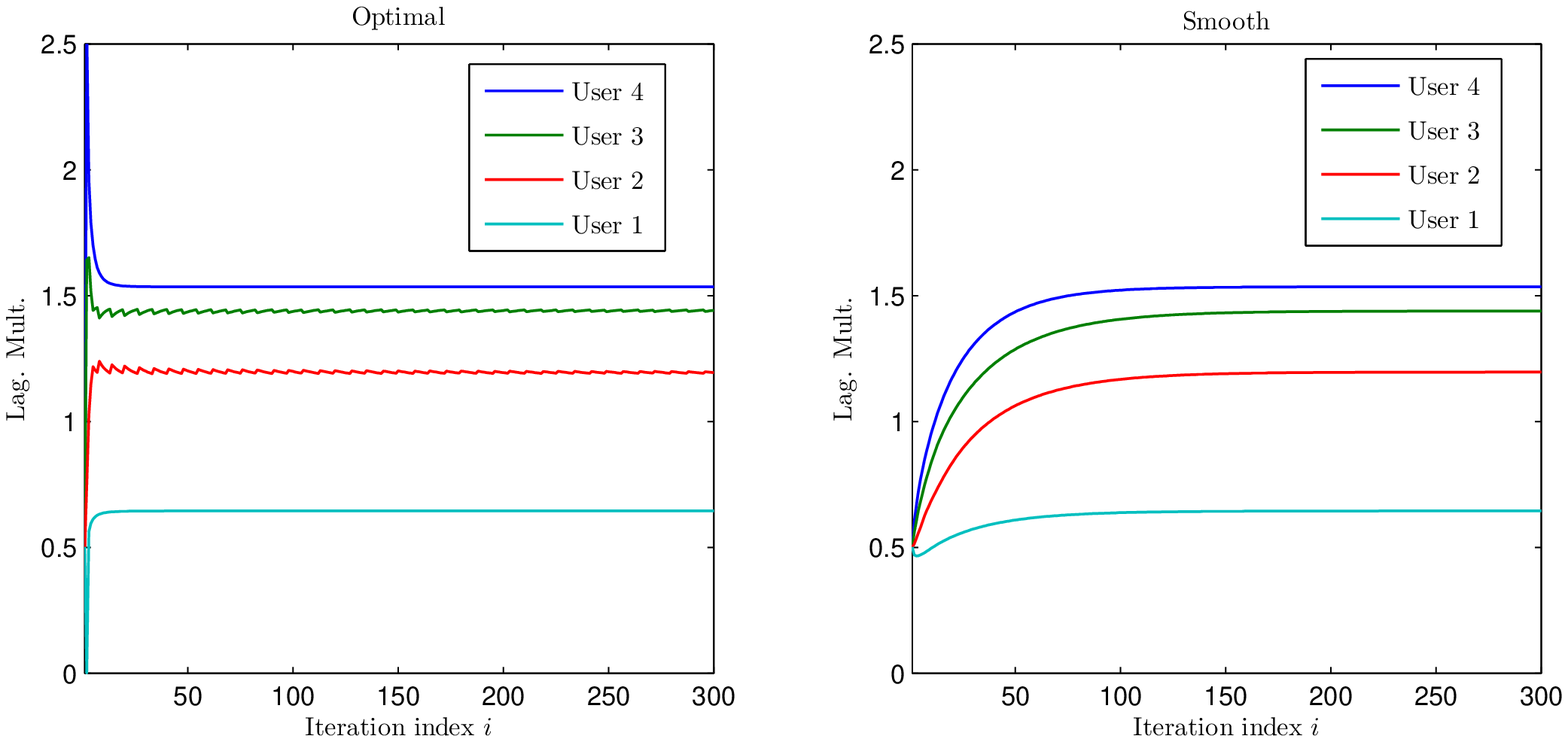}} \caption{Trajectories of the Lagrange Multipliers for off-line iterations. The iterations based on the optimal non-smooth policy (and decreasing stepsize) are shown in the left while the iterations based on the smooth policy (and constant stepsize) are shown in the right.} \label{F:Converg_Offline_Multip}
\end{figure}

To complement the analysis, we show in Figure \ref{F:Converg_Offline_Multip} the trajectories of the Lagrange multipliers. According to the analytical results, convergence occurs for both optimal iterations [cf. \eqref{E:original_off-line-iter}] and smooth iterations [cf. \eqref{E:off-line-iter}]. As explained in Section \ref{S:smooth_and_opt_Lag}, the hovering observed in Figure \ref{F:Converg_Offline_Rates_Powers} is due to the discontinuities of the optimal policy w.r.t. $\boldsymbol{\lambda}^{R}$. While Figure \ref{F:Converg_Offline_Multip} corroborates that the iterations in \eqref{E:original_off-line-iter} come closer and closer to the convergence point in the dual domain ($\boldsymbol{\lambda}^{R*}$), Figure \ref{F:Converg_Offline_Rates_Powers} illustrates that they fail to guarantee the same in the primal domain. On the other hand, the Lipschitz continuity of the smooth scheduling policy guarantees convergence in both dual and primal domains.

Based on both figures, it seems that in this specific case users 2 and 3 would have to share at least one channel. However, when they implement the optimum winner-takes-all scheduling, they keep competing to be the single winner of the channel. This competition ends only when the \emph{exact} value of $\boldsymbol{\lambda}^{R*}$ is found, but this only can be guaranteed after an infinite number of iterations.

The numerical tests reveal that the difference between the average power consumed by the smooth policy and the one by the optimum policy was $0.01$. This amount is considerably smaller than the bound $\varepsilon'=K\varepsilon=0.8$ given in Proposition \ref{P:converg}. As explained in footnote 4, such a bound is expected to be loose since it is derived for the worst-case scenario.

\begin{figure}
\def\epsfsize#1#2{0.62#1}
\centerline{\epsffile{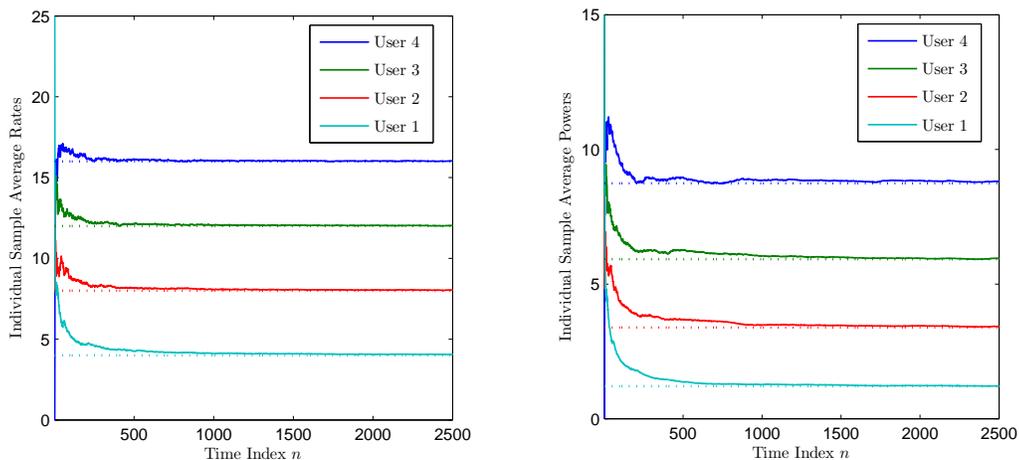}} \caption{Trajectories of the sample average rate (left) and sample average power (right) for online iterations. Ensemble values achieved by the off-line policy are represented as dotted lines.} \label{F:Converg_Online_Rates_Powers}
\end{figure}

\noindent \textbf{Test Case 2} (Convergence of the stochastic schemes):
The same set-up of Test Case 1 is used now to gauge convergence of the smooth stochastic schemes in \eqref{E:online-iter}. The left plot in Figure
\ref{F:Converg_Online_Rates_Powers} depicts the trajectories of the
sample average rate
$\hat{\bar{r}}_{m}[n]:=n^{-1}\sum_{q=1}^n\sum_{k=1}^K$ $[\mathbf{R}(\mathbf{J}[q],\hat{\boldsymbol{\lambda}}^R[q])]_{m,k}$ $[\mathbf{W}^s(\mathbf{J}[q],\hat{\boldsymbol{\lambda}}^R[q])]_{m,k}$ vs. the time index (online iterations) for every user, while the right plot depicts the corresponding trajectories of the sample average of the power $\hat{\bar{p}}_{m}[n]$. The figure illustrates not only that the stochastic schemes are able to achieve the same performance as the optimum off-line schemes (dotted line), but also that they converge within a few hundreds of iterations.

\begin{figure}
\def\epsfsize#1#2{0.62#1}
\centerline{\epsffile{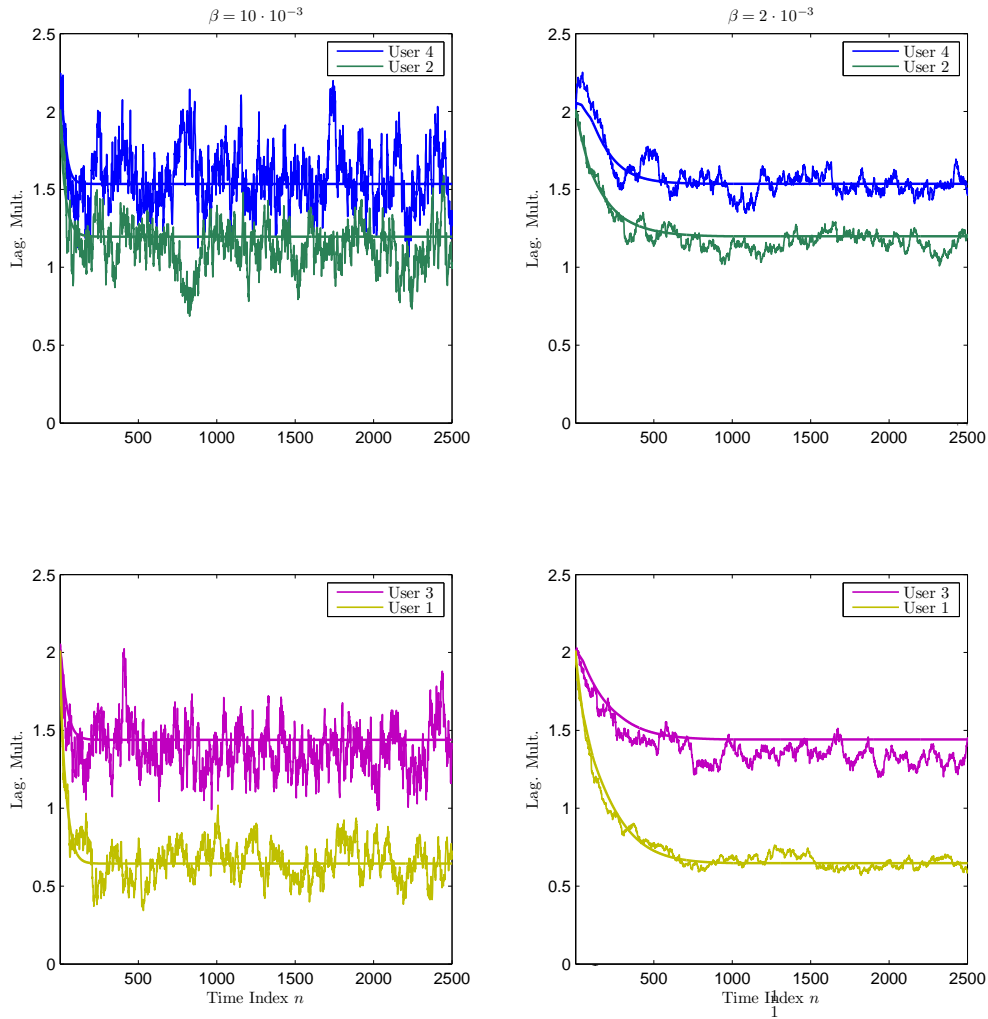}} \caption{Trajectories of estimated Lagrange multipliers $[\hat{\boldsymbol{\lambda}}^R[n]]_m$ for online iterations (solid lines). For comparison purposes, trajectories of the off-line iterations are also plotted (dotted lines).} \label{F:Converg_Online_Multip}
\end{figure}

To gain more insight about the behavior of the stochastic schemes, Figure \ref{F:Converg_Online_Multip} depicts the corresponding trajectories of the Lagrange multipliers $[\hat{\boldsymbol{\lambda}}^R[n]]_m$ for two different values of stepsize: $\beta=10\cdot10^{-3}$ (left column) and $\beta=2\cdot10^{-3}$ (right column). To facilitate visualization, trajectories of users 4 and 2 are shown in a different plot (top) from those of users 3 and 1 (bottom). For comparison purposes, the trajectories of the off-line iterations (with $i=n$) are also plotted using dotted lines. As Proposition \ref{P:online-locking} stated: (i) the trajectories of the online iterations remain locked to the trajectories of the off-line iterations; and, (ii) the smaller the step-size, the smaller the gap between online and off-line iterations.

\noindent \textbf{Test Case 3} (Performance comparison):
An OFDMA system was simulated here with $K=64$ subcarriers to serve $M=3$
users with $\mathbf{\check{r}}=[40, 70, 100]^T$ transmitting over a multi-path fading channel with eight taps and exponentially decaying gains. Figure \ref{F: Performance1} compares the \emph{overall} average transmit-power
for different SNR values. Results for five different resource allocation (RA)
policies are depicted: (i) the benchmark allocation obtained when P-CSI is
available (RA1) \cite{kn:xinGG_it08}; (ii) the optimum Q-CSIT based policy with the equally probable
channel quantizer of \cite[Sec. V-B]{kn:amfdgg06} (RA2); (iii) the smooth policy developed with the equally probable
channel quantizer of \cite[Sec. V-B]{kn:amfdgg06} (RA3); (iv) this paper's smooth policy
with a random quantizer (RA4); and (v) a policy based on Q-CSI which optimally adapts
$\mathbf{R}$ but fixes the channel scheduling matrix $\mathbf{W}$, and uses and on/off scheme for the power allocation $\mathbf{P}$. Not only the power consumption difference between (RA2) is (RA3) negligible, but their difference w.r.t. the optimum P-CSIT in (RA1) is small even for a (sub)-optimum channel quantizer. This is corroborated by the results for (RA4) that show that the power penalty for using a random quantizer is around 1dB. Finally, it is worth stressing the 6-8dB power savings of (RA3) relative to a heuristic scheme (RA5).

\begin{figure}
\def\epsfsize#1#2{0.62#1}
\centerline{\epsffile{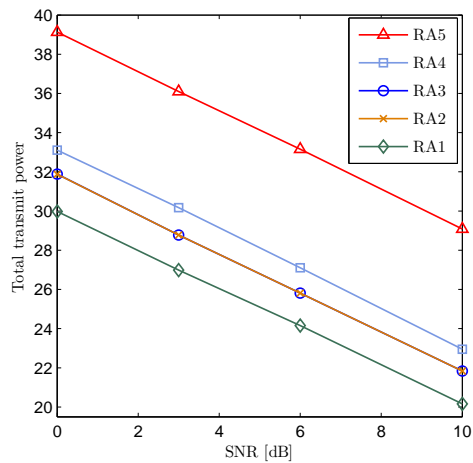}} \caption{Comparison of various resource
allocation schemes on the basis of average transmit-power [dB].} \label{F:
Performance1}
\end{figure}

Further numerical results assessing the performance of RA1,
RA3 and RA5 schemes over a wide range of parameter values
are summarized in Table \ref{T:MoreTestcases}. These
results confirm our previous conclusions, namely: (i) the near
optimality of R3, and (ii) the performance loss exhibited by
the heuristic schemes exemplified by R5. Results also show that when a more demanding set-up is simulated, the power savings due to the implementation of the optimum schemes are higher. This was expected because for easier scenarios (lower rate requirements, smaller number of users), ``reasonable'' heuristic policies can lead to a good solution.

\begin{table}
\caption{Total average weighted power for RA1,
RA3 and RA5 schemes. (Reference case: $K=64$, $M=3$,
$\mathbf{\check{r}}=[40,70,100]^T$,
SNR=6 $dB$; other cases describe
variation(s) w.r.t. the reference case.)} \label{T:MoreTestcases}
\centering
\begin{tabular}{|c|c|c|c|}
\hline
CASE & RA5 & RA3 & RA1 \\
\hline
Reference Case   & 29.9 & 21.7 & 19.9 \\
\hline
$[\mathbf{\check{r}}]_m=50$ & 22.6 & 18.3 & 16.2 \\
\hline
$[\mathbf{\check{r}}]_m=70$ & 26.8 & 21.7 & 19.6 \\
\hline
$K=128$ & 22.2  & 18.3 & 16.3 \\
\hline
$M=6$, $\mathbf{\check{r}}=[$40,52,64,76,88,100$]^T$ & 45.6  & 31.0  & 28.9 \\
\hline
$\Upsilon$ as in \eqref{E:p-r_ErgCap}  & 27.8  & 20.8  & 19.9 \\
\hline
\end{tabular}
\end{table}

\noindent \textbf{Test Case 4} (Sensitivity to the number of quantization regions): Table \ref{T:NumberRegions} lists the average transmit-power versus $L_k$ for a set-up with $M=3$ users and two different average rate requirements.
Consistent with orthogonal multiuser access based on Q-CSIT \cite{kn:amfdgg08}, \cite{kn:xwamgg2008}, the results in this table demonstrate that they lead
to a power loss no greater than 2-4 $dB$ w.r.t. the P-CSIT case
($L_{k}=\infty$) if $L>2$. (Recall
that for the simulated scenario, the lowest region will be inactive; hence,
$L=2$ implies one active region and one zero-rate/zero-power region.) Moreover, the resulting power gap shrinks as the
number of regions increases reaching a power loss of approximately
only 1 $dB$ with $L=8$ regions (3 feedback bits per channel).

\begin{table}
\caption{Total average weighted power for different values of the number of
regions per channel. (RA3 with $M=3$, $K=64$, and SNR=$6 dB$ $\forall m$ is implemented.)} \label{T:NumberRegions}
\begin{center}
\begin{tabular}{|c|c|c|c|c|c|c|c|}
\hline
\textbf{$\#$ of regions per channel } & 2   & 3  & 4  & 5  & 6  & 8  & $\infty$\\
\hline
\hline \textbf{Average Power [dB] if $\mathbf{\check{r}}=[50, 50, 50]^T$}         & 20.4 & 19.0 & 18.3 & 17.9 & 17.6 & 17.2 & 16.2 \\
\hline \textbf{Average Power [dB] if $\mathbf{\check{r}}=[40, 70, 100]^T$}       & 24.1 & 22.4 & 21.7 &  21.4 &  21.2 &  20.9 & 19.9\\
\hline
\end{tabular}
\end{center}
\end{table}

\section{Concluding Summary}\label{S:Conclusions}

This paper developed optimal scheduling and resource allocation policies for orthogonal multi-access transmissions over fading channels when both terminals and scheduler(s) have to rely only on quantized CSI. Focus has been placed on minimization of average power subject to average rate (capacity) constraints, but the results presented also when maximizing rate (capacity) subject to average power constraints.

Relative to systems with perfect CSI at the scheduler and channels with continuous fading, the main differences of the optimal policies show up in channel scheduling. It was shown that for most channel realizations the optimum scheduling amounts to a single (winner) user accessing the channel, while for a smaller set of realizations a few users share the resources. Optimal allocation in the sharing case is obtained as the solution of a linear program. This disjoint scheduling policy is also present in systems that exploit perfect CSI but operate over channels that are deterministic or have discrete fading distribution.

Having two different policies to schedule users not only incurs higher complexity relative
to the winner-takes-all case, but also complicates finding the optimum Lagrange multipliers needed to implement the optimal policies. To mitigate these challenges, a new scheduling scheme that combines the two different schedulers into a single one was developed. It was proved that this single scheme offers reduced complexity, facilitates finding the optimal Lagrange multipliers, and exhibits asymptotically optimal performance. Moreover, in order to facilitate practical implementation, stochastic schemes that do not need knowledge of the channel distribution, keep track of channel non-stationarities, reduce complexity and converge to the optimum solution were also developed. The last part of the paper was devoted to analyze the overhead associated to the novel schemes and present practical scenarios where the optimal policies derived can be implemented.\footnote{The views and conclusions contained in this document are those of the authors and should not be interpreted as representing the official policies, either expressed or implied, of the Army Research Laboratory or the U. S. Government.}

\section*{Appendix A:  Proof of Convexity of Eq. (\ref{E:optRA})}

If $\mathbf{x}$ collects all the optimization variables in
(\ref{E:optRA}), the convexity of (\ref{E:optRA}) can be
ensured if the cost function and all the constraints
satisfy $T_{x_i}^{f}:=\frac{\partial^2 f}{\partial x_i^2}\geq
0,~\forall i,$ and $T_{x_i,x_j}^{f}:=\frac{\partial^2 f}{\partial
x_i^2}\frac{\partial^2 f}{\partial x_j^2}-\left[\frac{\partial
f}{\partial x_i\partial x_j}\right]^2\geq 0,~\forall i,j$.
Since all
constraints are linear functions, both conditions
are satisfied $\forall~x_i,x_j$, and only the objective cost
function, $C$, must be checked. As the entries of
$\mathbf{\tilde{R}}$ are decoupled in $C$ (the cross-derivatives are zero) and the same happens with the entries
of $\mathbf{W}$. Hence, it suffices to consider three cases:
$T_{[\tilde{R}]_{m,k}}^{C}$, $T_{[\tilde{W}]_{m,k}}^{C}$, and
$T_{[\tilde{R}]_{m,k},[\tilde{W}]_{m,k}}^{C}$. The
second derivatives (after defining $r:=[\mathbf{\tilde{R}}(\mathbf{J})]_{m,k}$,
$w:=[\mathbf{W}(\mathbf{J})]_{m,k}$ for notational brevity) are:
\begin{eqnarray}
\label{E:partial2r}\frac{\partial^2 C}{\partial r^2}&=&\frac{\partial }{\partial
r}\left(\Dot{\Upsilon}\left(\frac{r}{w}\right)\right)={\ddot{\Upsilon}}\left(\frac{r}{w}\right)\frac{1}{w}\\
\label{E:partial2theta}
\frac{\partial^2 C}{\partial w^2}&=&\frac{\partial }{\partial
w}\left(\Dot{\Upsilon}\left(\frac{r}{w}\right)\frac{-r}{w}+\Upsilon\left(\frac{r}{w}\right)\right)={\ddot{\Upsilon}}\left(\frac{r}{w}\right)\frac{r^2}{w^3}\\
\label{E:partial2rtheta}
\frac{\partial^2 C}{\partial w \partial r}&=&\frac{\partial }{\partial
w}\left(\Dot{\Upsilon}\left(\frac{r}{w}\right)\right)={\ddot{\Upsilon}}\left(\frac{r}{w}\right)\frac{-r}{w^2}.
\end{eqnarray}

Expressions \eqref{E:partial2r}-\eqref{E:partial2rtheta} yield
$T_{[\tilde{R}]_{m,k},[\tilde{W}]_{m,k}}^{C}=0$, while both
$T_{[\tilde{R}]_{m,k}}^{C}\geq0$, and
$T_{[\tilde{W}]_{m,k}}^{C}\geq0$ provided that $\ddot{\Upsilon}\geq0$. Hence,
the problem in (\ref{E:optRA}) is convex if $\Upsilon$ is a convex
function.

\section*{Appendix B:  Proof of Proposition \ref{P:propChAll}}
Using \eqref{E:phi_chann_ind} and the fact that the multipliers must be non-negative, \eqref{E:kktt} and \eqref{E:kkttok} can be manipulated to yield
\begin{equation}\label{E:kkttphia}
\left([\mathbf{C}_W(\mathbf{J})]_{m,k}\Pr\{\mathbf{J}\} +
[\boldsymbol{\lambda}^{W*}(\mathbf{J})]_k\right)[\mathbf{W^{*}(J)}]_{m,k}=0,
~\forall m
\end{equation}
\begin{equation}\label{E:kkttphib}
[\boldsymbol{\alpha}^{W*}(\mathbf{J})]_{m,k}=([\mathbf{C}_W(\mathbf{J})]_{m,k}\Pr\{\mathbf{J}\}
+ [\boldsymbol{\lambda}^{W*}(\mathbf{J})]_k)\geq0, ~\forall m
\end{equation}
\begin{equation}\label{E:kkttphic}
[\boldsymbol{\lambda}^{W*}(\mathbf{J})]_k\geq0, ~\forall m.
\end{equation}
Slackness KKT
condition corresponding to the user-scheduling constraint also implies that
\begin{equation}\label{E:kkttphid}
[\boldsymbol{\lambda}^{W*}(\mathbf{J})]_k\left(\sum_{m=1}^M[\mathbf{W^{*}(J)}]_{m,k}-1\right)=0,~\forall k.
\end{equation}

Based on \eqref{E:kkttphia}-\eqref{E:kkttphid}, we have that:
\begin{enumerate}[{\it (i)}]
\item Since $m\in \mathcal{M}(\mathbf{J},k)$ requires the cost to be negative and minimum, we have to prove the validity of both. First, suppose $[\mathbf{W^{*}(J)}]_{m',k}>0$ for a user $m'$ whose cost $[\mathbf{C}_W(\mathbf{J})]_{m',k}$ is positive. Since
$[\boldsymbol{\lambda}^{W*}(\mathbf{J})]_k\geq0$, both factors $([\mathbf{C}_W(\mathbf{J})]_{m',k}\Pr\{\mathbf{J}\} +
[\boldsymbol{\lambda}^{W*}(\mathbf{J})]_k)$ and $[\mathbf{W^{*}(J)}]_{m',k}>0$ in (\ref{E:kkttphia}) are positive, which contradicts the equality required by (\ref{E:kkttphia}).
Suppose now $[\mathbf{W^{*}(J)}]_{m',k}>0$ for a user
$m'$ such that $[\mathbf{C}_W(\mathbf{J})]_{m',k}>[\mathbf{c}_W^*(\mathbf{J},k)]_k$. Then, satisfaction of (\ref{E:kkttphia}) for user $m'$ requires
$[\boldsymbol{\lambda}^{W*}(\mathbf{J})]_k=-[\mathbf{C}_W(\mathbf{J})]_{m,k'}\Pr\{\mathbf{J}\}$. Substituting this value into (\ref{E:kkttphib}) to obtain the multiplier for a user $m_k\in\mathcal{M}(\mathbf{J},k)$ yields
$[\boldsymbol{\alpha}^{W*}(\mathbf{J})]_{m_k,k}=[\mathbf{c}_W^*(\mathbf{J},k)]_k\Pr\{\mathbf{J}\}-[\mathbf{C}_W(\mathbf{J})]_{m',k}\Pr\{\mathbf{J}\}$, which is a negative number and hence contradicts the right hand side of (\ref{E:kkttphib}).
\item If $|\mathcal{M}(\mathbf{J},k)|>0$, then $[\mathbf{C}_W(\mathbf{J})]_{m,k}<0$ for $m \in \mathcal{M}(\mathbf{J},k)$. This requires
$[\boldsymbol{\lambda}^{W*}(\mathbf{J})]_k>0$ in (\ref{E:kkttphib}). Substituting the latter into (\ref{E:kkttphid}), the statement follows.
\item By construction, $|\mathcal{M}(\mathbf{J},k)|=0$ if and only if $[\mathbf{C}_W(\mathbf{J})]_{m,k}>0~\forall m$. This implies that
if $|\mathcal{M}(\mathbf{J},k)|=0$, then (\ref{E:kkttphib}) will be strictly positive $\forall m$, and thus (\ref{E:kkttphia}) can be only hold if $[\mathbf{W^{*}(J)}]_{m,k'}=0$ $\forall m$.
\end{enumerate}

\section*{Appendix C:  Proof of Lemma \ref{L:propDSmoth}}
To prove the first part of the lemma, re-write the Lagrangian in (\ref{E:simpl_Lagrangian}) using the cost in (\ref{E:phi_chann_ind}) as
\begin{equation}
\mathcal{L}(\boldsymbol{\lambda}^R,\mathbf{\tilde{R}}(\mathbf{J}),\mathbf{W}(\mathbf{J}))=
\sum_{\forall \mathbf{J} \in \mathcal{J}}\left(\sum_{k=1}^{K}\sum_{m=1}^{M}[\mathbf{C}_W(\mathbf{J}, \boldsymbol{\lambda}^R)]_{m,k}[\mathbf{W}(\mathbf{J})_{m,k}]\right)\Pr\{\mathbf{J}\}
+\sum_{m=1}^M [\boldsymbol{\lambda}^R]_m[\mathbf{\check{r}}]_m.
\end{equation}
The dual function can be written as
\begin{equation}
D(\boldsymbol{\lambda}^R)=
\sum_{\forall \mathbf{J} \in \mathcal{J}}\left(\sum_{k=1}^{K}[\mathbf{c}_W^*(\mathbf{J},\boldsymbol{\lambda}^R)]_{k}[\mathbf{W}^*(\mathbf{J})]_{m^*,k}\right)\Pr\{\mathbf{J}\}
+\sum_{m=1}^M [\boldsymbol{\lambda}^R]_m[\mathbf{\check{r}}]_m
\end{equation}
and the smooth version of the dual function as
\begin{equation}
D^s(\boldsymbol{\lambda}^R)=
\sum_{\forall \mathbf{J} \in \mathcal{J}}\left(\sum_{k=1}^{K}\sum_{m\in\mathcal{M}(\mathbf{J},k)}[\mathbf{C}_W(\mathbf{J},\boldsymbol{\lambda}^R)]_{m,k}[\mathbf{W}^s(\mathbf{J})]_{m,k}\right)\Pr\{\mathbf{J}\}
+\sum_{m=1}^M [\boldsymbol{\lambda}^R]_m[\mathbf{\check{r}}]_m.
\end{equation}
 Based on the definition of $\mathcal{M}(\mathbf{J},k)$ and Proposition \ref{P:propWSmoth}, it follows that $[\mathbf{W}^*(\mathbf{J})]_{m^*,k}=\sum_{m\in\mathcal{M}(\mathbf{J},k)}[\mathbf{W}^s(\mathbf{J})]_{m,k}$ $\forall k$. Using this equality, consider the difference
\begin{equation}\label{E:diff_bet_Ds_D_App}
D^s(\boldsymbol{\lambda}^R)-D(\boldsymbol{\lambda}^R)=\sum_{\forall \mathbf{J} \in \mathcal{J}}\sum_{k=1}^{K}\left(\sum_{m\in\mathcal{M}(\mathbf{J},k)}\left([\mathbf{C}_W
(\mathbf{J},\boldsymbol{\lambda}^R)]_{m,k}-\vphantom{\frac{1^^1}{1^^1}}[\mathbf{c}_W^*(\mathbf{J},\boldsymbol{\lambda}^R)]_{k}\right)[\mathbf{W}^s(\mathbf{J})]_{m,k}
\right)\Pr\{\mathbf{J}\}.
\end{equation}

It holds by construction that $[\mathbf{C}_W(\mathbf{J},\boldsymbol{\lambda}^R)]_{m,k}-[\mathbf{c}_W^*(\mathbf{J},\boldsymbol{\lambda}^R)]_{k}\geq0$ and $[\mathbf{C}_W(\mathbf{J},\boldsymbol{\lambda}^R)]_{m,k}-[\mathbf{c}_W^*(\mathbf{J},\boldsymbol{\lambda}^R)]_{k}<\varepsilon$. Substituting these expressions into \eqref{E:diff_bet_Ds_D_App} yields, respectively,
\begin{equation}\label{E:1st_cond_Prop4_Ap}
D^s(\boldsymbol{\lambda}^R)-D(\boldsymbol{\lambda}^R)\geq0
\end{equation}
\begin{equation}\label{E:2nd_cond_Prop4_Ap}
D^s(\boldsymbol{\lambda}^R)-D(\boldsymbol{\lambda}^R)<\sum_{\forall \mathbf{J} \in \mathcal{J}}\sum_{k=1}^{K}\sum_{m\in\mathcal{M}(\mathbf{J},k)}\varepsilon[\mathbf{W}^s(\mathbf{J})]_{m,k}\Pr\{
\mathbf{J}\}\leq \sum_{\forall \mathbf{J} \in \mathcal{J}}\sum_{k=1}^{K}\varepsilon\Pr\{
\mathbf{J}\}=K\varepsilon
\end{equation}
where in \eqref{E:1st_cond_Prop4_Ap} we have used that $[\mathbf{W}^s(\mathbf{J})]_{m,k}\geq0$ and in \eqref{E:2nd_cond_Prop4_Ap} we have used that $\sum_{m\in\mathcal{M}(\mathbf{J},k)}[\mathbf{W}^s(\mathbf{J})]_{m,k}\leq 1$. Equations \eqref{E:1st_cond_Prop4_Ap} and \eqref{E:2nd_cond_Prop4_Ap} prove part (i) of Lemma \ref{L:propDSmoth}.

To establish part (ii), since $[\partial^s D(\boldsymbol{\lambda}^R)]_m$ can be written as a summation of $[\mathbf{R}^*(\mathbf{J},\boldsymbol{\lambda}^R)]_{m,k}[\mathbf{W}^s(\mathbf{J},\boldsymbol{\lambda}^R)]_{m,k}$ terms, we will show that $[\partial^s D(\boldsymbol{\lambda}^R)]_m$ is Lipschitz continuous w.r.t. $\boldsymbol{\lambda}^R$ by arguing that both $\mathbf{W}^s(\mathbf{J},\boldsymbol{\lambda}^R)$ and $\mathbf{R}^*(\mathbf{J},\boldsymbol{\lambda}^R)$ are Lipschitz continuous w.r.t. $\boldsymbol{\lambda}^R$. On the one hand, continuity of $\mathbf{W}^s(\mathbf{J},\boldsymbol{\lambda}^R)$ is ensured by Proposition \ref{P:propWSmoth}-(iii). Obtaining the Lipschitz constant for this case is trivial, because $[\mathbf{W}^s(\mathbf{J},\boldsymbol{\lambda}^R)]_{m,k}$ is differentiable by construction [cf. \eqref{E:WSmoth}]. On the other hand, since $[\mathbf{R}^*(\mathbf{J},\boldsymbol{\lambda}^R)]_{m,k}$ depends only on the $m$th entry of $\boldsymbol{\lambda}^R$ [cf. Proposition \ref{P:propRAll}], it suffices to consider how $[\mathbf{R}^*(\mathbf{J})]_{m,k}$ varies with $[\boldsymbol{\lambda}^R]_m$. Since $\Upsilon$ is strictly convex, it is easy to deduce that $\dot{\Upsilon}$ is a continuous monotonic one-to-one function, and so is $\dot{\Upsilon}^{-1}$. While continuity of $\dot{\Upsilon}^{-1}$ implies continuity of $[\mathbf{R}^*(\mathbf{J},\boldsymbol{\lambda}^R)]_{m,k}$ w.r.t. $[\boldsymbol{\lambda}^R]_m$ [cf. \eqref{E:rateopt_nom}], its monotonicity together with the fact that the rate is bounded, gives the Lipschitz property.

\section*{Appendix D: Properties of the Updating Matrices}
This appendix analyzes the behavior of the smooth subgradient in Lemma \ref{L:propDSmoth}. The main result is summarized in Lemma \ref{L:PD_of_Jacobian}, which is critical for proving convergence of both the off-line iterations in Proposition \ref{P:converg} and the online iterations in Proposition \ref{P:online-locking}.

Define $\mathbf{f}^{av}$ and $\mathbf{f}$ as $M\times 1$ \emph{vector valued functions} with entries
\begin{equation}
[\mathbf{f}(\mathbf{J},\boldsymbol{\lambda}^R)]_m:=[\check{\mathbf{r}}]_m-\sum_{\forall
k}[\mathbf{R}^*(\mathbf{J},\boldsymbol{\lambda}^R)]_{m,k}[\mathbf{W}^s(\mathbf{J},\boldsymbol{\lambda}^R)]_{m,k}
\end{equation}
\begin{equation}
[\mathbf{f}^{av}(\boldsymbol{\lambda}^R)]_m:=[\check{\mathbf{r}}]_m-\sum_{\forall
\mathbf{J}}\sum_{\forall
k}[\mathbf{R}^*(\mathbf{J},\boldsymbol{\lambda}^R)]_{m,k}[\mathbf{W}^s(\mathbf{J},\boldsymbol{\lambda}^R)]_{m,k}\Pr\{\mathbf{J}\}=\sum_{\forall
\mathbf{J}}[\mathbf{f}(\mathbf{J},\boldsymbol{\lambda}^R)]_m\Pr\{\mathbf{J}\}
\end{equation}
which coincide with the instantaneous and average smooth subgradients $\partial^s D^s(\boldsymbol{\lambda}^R,n)$ (Section \ref{S:stoch}) and $\partial^s D(\boldsymbol{\lambda}^R)$ (Section \ref{S:smooth_and_opt_Lag}), respectively.

The Jacobian $M\times M$ matrices of those functions are
$[\mathbf{\Delta}^s(\mathbf{J})]_{q,m}={\partial [\mathbf{f}(\mathbf{J},\boldsymbol{\lambda}^R)]_q}/{\partial [\boldsymbol{\lambda}^R]_m}$ and $[\mathbf{\Delta}^s]_{q,m}=\sum_{\forall
\mathbf{J}}[\mathbf{\Delta}^s(\mathbf{J})]_{q,m}\Pr\{\mathbf{J}\}$,
respectively. Since the entries of $\mathbf{f}$ depend on $\mathbf{R}^*$ and $\mathbf{W}^s$, it follows that
\begin{eqnarray}
\mathbf{\Delta}^s(\mathbf{J})&:=&-\left(\mathbf{\Delta}_R^s(\mathbf{J})+\mathbf{\Delta}_W^s(\mathbf{J})\right),~\mathrm{where}~\label{E:Jac_sum_jac_R_W}\\
&~&[\mathbf{\Delta}_R^s(\mathbf{J})]_{q,m}
:=\sum_{\forall k}[\mathbf{W}^s(\mathbf{J},\boldsymbol{\lambda}^R)]_{q,k}{\partial[\mathbf{R}^*(\mathbf{J},\boldsymbol{\lambda}^R)]_{q,k}}/{\partial [\boldsymbol{\lambda}^R]_m}~\mathrm{and}~\label{E:JacR}\\
&~&[\mathbf{\Delta}_W^s(\mathbf{J})]_{q,m}:=\sum_{\forall k}[\mathbf{R}^*(\mathbf{J},\boldsymbol{\lambda}^R)]_{q,k} {\partial[\mathbf{W}^s(\mathbf{J},\boldsymbol{\lambda}^R)]_{q,k}}/{\partial [\boldsymbol{\lambda}^R]_m}.\label{E:JacW}
\end{eqnarray}

\begin{lemma}\label{L:PD_of_Jacobian}
\emph{Matrices $\mathbf{\Delta}^s(\mathbf{J})$ and $\mathbf{\Delta}^s$ are: (i) negative definite, and (ii) with bounded eigenvalues.}
\end{lemma}
\begin{IEEEproof}
Since $\mathbf{\Delta}^s$ is a weighted sum of $\mathbf{\Delta}^s(\mathbf{J})$, it suffices to prove (i) and (ii) for $\mathbf{\Delta}^s(\mathbf{J})$. To simplify notation, consider a single channel and drop the subindex $k$ (extension for $K>1$ is straightforward). To prove (i), we will show first that $\mathbf{\Delta}_R^s(\mathbf{J})$ is positive definite (PD), and then that $\mathbf{\Delta}_W^s(\mathbf{J})$ is semi-PD (SPD); thus, the sum of both is PD and $\mathbf{\Delta}^s(\mathbf{J})$ is negative definite.

Clearly, the derivative of the rate in \eqref{E:rateopt_nom} is zero if $q\neq m$; hence, $\mathbf{\Delta}_R^s(\mathbf{J})$ is diagonal. Using the theorem of the inverse function, the diagonal entries are
\begin{equation}\label{E:Hessian_Rate}
[\mathbf{\Delta}_R^s(\mathbf{J})]_{m,m}=\frac{1}{\ddot{\Upsilon}([\mathbf{R}^*(\mathbf{J},\boldsymbol{\lambda}^R)]_{m})}\frac{1}{[\boldsymbol{\mu}]_m},\;\;\forall m.
\end{equation}
Since $\Upsilon$ is assumed strictly convex and the rate is bounded, the diagonal elements in \eqref{E:Hessian_Rate} are finite, positive and nonzero; thus, $\mathbf{\Delta}_R^s(\mathbf{J})$ is PD.

To prove that $\mathbf{\Delta}_W^s(\mathbf{J})$ is SPD, define first $\mathbf{D}_R(\mathbf{J})$ as a $M\times M$ diagonal matrix with entries $[\mathbf{D}_R(\mathbf{J})]_{m,m}$ $:=[\mathbf{R}^*(\mathbf{J},\boldsymbol{\lambda}^R)]_{m}$, and $\mathbf{\Delta}_C^s(\mathbf{J}) $ with entries $[\mathbf{\Delta}_C^s(\mathbf{J}) ]_{q,m}:=-\partial [\mathbf{W}^s(\mathbf{J},\boldsymbol{\lambda}^R)]_{q}$ $/\partial [\mathbf{C}_W(\mathbf{J},\boldsymbol{\lambda}^R)]_{m}$. Since $\mathbf{W}^s(\mathbf{J},\boldsymbol{\lambda}^R)$ can be also written as a function of $\mathbf{C}_W(\mathbf{J},\boldsymbol{\lambda}^R)$ [cf. \eqref{E:WSmoth}], $\mathbf{\Delta}_C^s(\mathbf{J})$ represents the Jacobian matrix of the vector function $[[\mathbf{W}^s(\mathbf{J},\boldsymbol{\lambda}^R)]_1,\ldots,[\mathbf{W}^s(\mathbf{J},\boldsymbol{\lambda}^R)]_M]$ w.r.t. the vector variable $-[[\mathbf{C}_W(\mathbf{J},\boldsymbol{\lambda}^R)]_1,$ $\ldots,[\mathbf{C}_W(\mathbf{J},\boldsymbol{\lambda}^R)]_M]$. Based on the previous definitions, $\mathbf{\Delta}_W^s(\mathbf{J})$ can be written as
\begin{equation}
\mathbf{\Delta}_W^s(\mathbf{J}):=\mathbf{D}_R(\mathbf{J})  \mathbf{\Delta}_C^s(\mathbf{J}) \mathbf{D}_R(\mathbf{J}).
\end{equation}
The multiplication from the left corresponds to the rate product in the definition of $\mathbf{\Delta}_W^s(\mathbf{J})$ in \eqref{E:JacW}, while the multiplication from the right represents the derivative of $-\mathbf{C}_W(\mathbf{J},\boldsymbol{\lambda}^R)$ w.r.t. $\boldsymbol{\lambda}^R$  (chain rule). Since the product of SPD matrices of the form $\mathbf{X}\times \mathbf{Y} \times \mathbf{X}$ is SPD if both $\mathbf{X}$ and $ \mathbf{Y}$ are SPD, and $\mathbf{D}_R(\mathbf{J})$ is PD (diagonal matrix with positive entries), it suffices to show that $\mathbf{\Delta}_C^s(\mathbf{J})$ is SPD.

To find entries of $\mathbf{\Delta}_C^s(\mathbf{J})$ four different cases have to be considered: (i) $q\notin\mathcal{M}^s(\mathbf{J})$; (ii) $q\in\mathcal{M}^s(\mathbf{J})$ and $|\mathcal{M}^s(\mathbf{J})|=1$; (iii) $q\in\mathcal{M}^s(\mathbf{J})$, $|\mathcal{M}^s(\mathbf{J})|>1$ and $[\mathbf{C}_W(\mathbf{J},\boldsymbol{\lambda}^R)]_{m}>[\mathbf{c}_W^*(\mathbf{J},
\boldsymbol{\lambda}^R)]$; and (iv) $q\in\mathcal{M}^s(\mathbf{J})$, $|\mathcal{M}^s(\mathbf{J})|>1$ and $[\mathbf{C}_W(\mathbf{J},\boldsymbol{\lambda}^R)]_{m}=[\mathbf{c}_W^*(\mathbf{J},
\boldsymbol{\lambda}^R)]$. For the two first cases, $[\mathbf{W}^s(\mathbf{J}^s)]_{m}$ is constant and therefore its derivative is zero. The expressions for the derivatives of (iii) and (iv) are given in \eqref{E:wsder_non_win_l} and \eqref{E:wsder_win_l}, respectively. Those have been obtained after manipulating \eqref{E:WSmoth} and defining $n_{m}:=1-\left([\mathbf{C}_W(\mathbf{J},\boldsymbol{\lambda}^R)]_{q}
        -[\mathbf{c}_W^*(\mathbf{J},\boldsymbol{\lambda}^R)]\right)/\varepsilon$ and $d:=\sum_{m'\in\mathcal{M}^s(\mathbf{J},k)}n_{m'}^2$ (recall that $n_m\in[0,1]$ and $n_{m^*}=1$).

\begin{subequations}
\label{E:wsder_non_win_l}
\begin{alignat}{1}
\label{E:wsder_non_win_l1}
[\mathbf{\Delta}_C^s(\mathbf{J})]_{m,m}&=\frac{2}{\varepsilon} \frac{n_{m}\sum_{\substack{m'\in\mathcal{M}^s(\mathbf{J})\\ m'\neq m}}n_{m'}^2}{d^2},\;\:m\neq m^*\hspace{4.52cm}\\
\label{E:wsder_non_win_l2}
[\mathbf{\Delta}_C^s(\mathbf{J})]_{q,m}&=-\frac{2}{\varepsilon} \frac{n_{q}^2n_{m}}{d^2},\;\:m\neq m^*\hspace{4.52cm}
\end{alignat}
\end{subequations}
\begin{subequations}
\label{E:wsder_win_l}
\begin{alignat}{1}
\label{E:wsder_win_l1}
[\mathbf{\Delta}_C^s(\mathbf{J})]_{m^*,m^*}&=\frac{2}{\varepsilon} \frac{\sum_{\substack{m'\in\mathcal{M}^s(\mathbf{J})\\ m'\neq m^*}}n_{m'}}{d^2},\;\:m=m^*\\
\label{E:wsder_win_l2}
[\mathbf{\Delta}_C^s(\mathbf{J})]_{q,m^*}&=-\frac{2}{\varepsilon} \frac{n_{q}+n_{q}\sum_{\substack{m'\in\mathcal{M}^s(\mathbf{J})\\ m'\neq m^*}}n_{m'}^2-n_{q}^2\sum_{\substack{m'\in\mathcal{M}^s(\mathbf{J})\\ m'\neq m^*}}n_{m'}}{d^2},\;\:m= m^*
\end{alignat}
\end{subequations}

Matrix $\mathbf{\Delta}_C^s(\mathbf{J})$ has several useful properties, namely: (i) it has zero column sum; (ii) it has zero row sum; (iii) all diagonal entries are positive; and (iv) for columns $m \neq m^*$, all non-diagonal entries are non-positive. Using \eqref{E:wsder_non_win_l} and \eqref{E:wsder_win_l} and these properties, the following result can be established to prove that $\mathbf{\Delta}_W^s(\mathbf{J})$ is SPD and thus conclude the proof of Lemma \ref{L:PD_of_Jacobian}-(i).
\begin{lemma}\label{L:proof_Lemma}
\emph{It holds for $\mathbf{\Delta}_C^s(\mathbf{J})$ that: (i) it has one zero eigenvalue; and, (ii) it is SPD.}
\end{lemma}
\begin{IEEEproof}
Proving Lemma \ref{L:proof_Lemma}-(i) only requires considering the products $\mathbf{1}^T\mathbf{\Delta}_C^s(\mathbf{J})$ and $\mathbf{\Delta}_C^s(\mathbf{J})\mathbf{1}$, where $\mathbf{1}$ is the $M\times 1$ all-ones vector. Since $\mathbf{\Delta}_C^s(\mathbf{J})$ has zero-column and zero-row sums,  $\mathbf{1}^T\mathbf{\Delta}_C^s(\mathbf{J})=\mathbf{\Delta}_C^s(\mathbf{J})\mathbf{1}=\mathbf{0}$. This implies that $\mathbf{1}$ is both a left and a right eigenvector of $\mathbf{\Delta}_C^s(\mathbf{J})$ whose associated eigenvalue is $0$. The proof of (ii) relies on the structure of $\mathbf{\Delta}_C^s(\mathbf{J})$. According to \eqref{E:wsder_non_win_l} and \eqref{E:wsder_win_l}, all rows and columns of $\mathbf{\Delta}_C^s(\mathbf{J})$ except $m^*$ have a regular structure. Consider an $M\times M$ matrix $\mathbf{U}$ such that $[\mathbf{U}]_{m,m}:=1$ $\forall m$, $[\mathbf{U}]_{m^*,m}:=1$ $\forall m$; and $[\mathbf{U}]_{m,m'}:=0$, otherwise. It is clear that $\mathbf{U}$ has rank $M$ and the  range of $\mathbf{U}^T$ is $\mathbb{R}^M$. Consider now the matrix $\mathbf{V}(\mathbf{J}):=\mathbf{U}\times\mathbf{\Delta}_C^s(\mathbf{J})\times\mathbf{U}^T$. Due to the structure of $\mathbf{U}$ and $\mathbf{\Delta}_C^s(\mathbf{J})$, it follows that $[\mathbf{V}(\mathbf{J})]_{m,m'}=0$ if either $m=m^*$ or $m'=m^*$, while $[\mathbf{V}(\mathbf{J})]_{m,m'}=[\mathbf{\Delta}_C^s(\mathbf{J})]_{m,m'}$. In words, $\mathbf{V}(\mathbf{J})$ is a copy of $\mathbf{\Delta}_C^s(\mathbf{J})$ were both the $m^*$th column and the $m^*$th row have been set to zero. Suppose now that $\mathbf{V}(\mathbf{J})$ is SPD, meaning that $\mathbf{\tilde{x}}^T\mathbf{V}(\mathbf{J})\mathbf{\tilde{x}}\geq 0$ $\forall \mathbf{\tilde{x}}\in\mathbb{R}^M$ or equivalently $\mathbf{\tilde{x}}^T\mathbf{U}\times\mathbf{\Delta}_C^s(\mathbf{J})\times\mathbf{U}^T\mathbf{\tilde{x}}\geq0$. Setting $\mathbf{x}=\mathbf{U}^T\mathbf{\tilde{x}}$, we can conclude that $\mathbf{x}^T\mathbf{\Delta}_C^s(\mathbf{J})\mathbf{x}\geq 0$, and therefore $\mathbf{\Delta}_C^s(\mathbf{J})$ is SPD. The next lemma establishes that $\mathbf{V}(\mathbf{J})$ is in fact SPD and hence $\mathbf{\Delta}_C^s(\mathbf{J})$ is SPD, as asserted by Lemma \ref{L:proof_Lemma}-(ii).
\end{IEEEproof}
\begin{lemma}\label{L:proof_Lemma_SPD_wo_opt_rowcol}
\emph{It holds for $\mathbf{V}(\mathbf{J})$ that: (i) it has one zero eigenvalue; and, (ii) it is SPD.}
\end{lemma}
\begin{IEEEproof}
Without loss of generality, assume that $m^*=M$ and define $\mathbf{Q}(\mathbf{J})$ as the $(M-1)\times (M-1)$ matrix whose $m$th column is formed by the $M-1$ first entries of the $m$th column of $\mathbf{V}(\mathbf{J})$; i.e., the all-zero column and all-zero row corresponding to the optimum user have been dropped. It is clear that the eigenvalues of $\mathbf{V}(\mathbf{J})$ are all the eigenvalues of $\mathbf{Q}(\mathbf{J})$ plus a zero eigenvalue. Hence, in order to prove Lemma \ref{L:proof_Lemma_SPD_wo_opt_rowcol}, it suffices to show that $\mathbf{Q}(\mathbf{J})$ is PD.

To prove that $\mathbf{Q}(\mathbf{J})$ is PD, let $\mathbf{D}(\mathbf{J})_N$ denote an $(M-1)\times (M-1)$ diagonal matrix with positive entries $[\mathbf{D}(\mathbf{J})_N]_{m,m}=n_m$ and recall that $\mathbf{I}_{M-1}$ and $\mathbf{1}_{M-1,M-1}$ denote the identity and all-ones $(M-1)\times (M-1)$ matrices, respectively. Using this notation, \eqref{E:wsder_non_win_l} can be written in matrix form as
\begin{equation}\label{E:Q_in_Lemma_is_PD}
\mathbf{Q}(\mathbf{J})=\frac{2}{\varepsilon d^2}\mathbf{D}(\mathbf{J})_N [\mathbf{I}_{M-1} + \mathbf{\Delta}_N(\mathbf{J})]
\end{equation}
where
\begin{equation}
\mathbf{\Delta}_N(\mathbf{J})=\mathrm{Tr}(\mathbf{D}_N(\mathbf{J})\mathbf{D}_N(\mathbf{J}))\mathbf{I}_{M-1}-\mathbf{D}_N(\mathbf{J})\mathbf{1}_{M-1,M-1}\mathbf{D}_N(\mathbf{J}).
\end{equation}
Matrix $\mathbf{\Delta}_N(\mathbf{J})$ is SPD because all its eigenvalues are nonnegative. In fact, it is easy to see that the eigenvalues of $\mathbf{\Delta}_N(\mathbf{J})$ are 0 and $\mathrm{Tr}(\mathbf{D}_N(\mathbf{J})\mathbf{D}_N(\mathbf{J}))$, the latter one with multiplicity $M-2$. This property implies that the factor $\mathbf{I}_{M-1} + \mathbf{\Delta}_N(\mathbf{J})$ in \eqref{E:Q_in_Lemma_is_PD} is PD. Since $2/\varepsilon d^2>0$ and $\mathbf{D}(\mathbf{J})_N$ in \eqref{E:Q_in_Lemma_is_PD} is also PD (diagonal with positive entries), it follows that $\mathbf{Q}(\mathbf{J})$ is PD, concluding the proof of Lemma \ref{L:proof_Lemma_SPD_wo_opt_rowcol}.
\end{IEEEproof}

Summarizing, we have proved that $\mathbf{\Delta}^s(\mathbf{J})$ is PD because it can be written as $\mathbf{\Delta}^s(\mathbf{J})=\mathbf{\Delta}_R^s(\mathbf{J})+\mathbf{\Delta}_W^s(\mathbf{J})$, where $\mathbf{\Delta}_R^s(\mathbf{J})$ is a PD and $\mathbf{\Delta}_W^s(\mathbf{J})$ is SPD. Matrix $\mathbf{\Delta}_R^s(\mathbf{J})$ is PD because it is diagonal with positive entries [cf. \eqref{E:Hessian_Rate}]. On the other hand, $\mathbf{\Delta}_W^s(\mathbf{J})$ is SPD because it can be written as
$\mathbf{D}_R(\mathbf{J}) \mathbf{\Delta}_C^s(\mathbf{J}) \mathbf{D}_R(\mathbf{J})$, where $\mathbf{D}_R(\mathbf{J})$ is PD (diagonal with positive entries) and $\mathbf{\Delta}_C^s(\mathbf{J})$ is SPD [cf. Lemmas \ref{L:proof_Lemma} and \ref{L:proof_Lemma_SPD_wo_opt_rowcol}].

To show Lemma \ref{L:PD_of_Jacobian}-(ii) we only have to show that the eigenvalues of $\mathbf{\Delta}^s(\mathbf{J})$ are bounded. This follows from the fact that the entries of both $\mathbf{\Delta}_R^s(\mathbf{J})$ and $\mathbf{\Delta}_W^s(\mathbf{J})$ are bounded. Specifically, the strict convexity of $\Upsilon$ guarantees that the non-zero entries of $\mathbf{\Delta}_R^s(\mathbf{J})$ are finite [cf. the denominator in \eqref{E:Hessian_Rate}]. In addition, the absolute value of the entries of $\mathbf{\Delta}_C^s(\mathbf{J})$ in \eqref{E:wsder_non_win_l1}, \eqref{E:wsder_non_win_l2}, \eqref{E:wsder_win_l1}, and \eqref{E:wsder_win_l2} can be safely upper bounded by $1/\varepsilon$, $1/\varepsilon$, $2(M-1)/\varepsilon$, and $(M-1)/\varepsilon$, respectively.

\end{IEEEproof}

\section*{Appendix E: Proof of Proposition \ref{P:converg}-(ii)}
Since Proposition \ref{P:converg}-(ii) provides upper and lower bounds for $D^s(\boldsymbol{\lambda}^{Rs})$, we will prove each separately. Recall that $\boldsymbol{\lambda}^{Rs}$ denotes the limit of the $\varepsilon'$-subgradient iteration and $\boldsymbol{\lambda}^{R*}$ the optimal solution of \eqref{E:dual_problem}.

To prove the upper bound, we rely on Lemma \ref{L:propDSmoth}-(i) which ensures that  $D^s(\boldsymbol{\lambda}^{R})<D(\boldsymbol{\lambda}^{R})+\varepsilon'$ $\forall \boldsymbol{\lambda}^{R}$. Substituting $\boldsymbol{\lambda}^{R}=\boldsymbol{\lambda}^{Rs}$ into the last inequality yields
\begin{equation}\label{E:2ndCondProofApE}
D^s(\boldsymbol{\lambda}^{Rs})< D(\boldsymbol{\lambda}^{Rs})+\varepsilon'.
\end{equation}
Moreover, since $\boldsymbol{\lambda}^{R*}$ is the value maximizing $D(\boldsymbol{\lambda}^{R})$, it holds that $D(\boldsymbol{\lambda}^{Rs})\leq D(\boldsymbol{\lambda}^{R*})$. Substituting this condition into \eqref{E:2ndCondProofApE} one can readily obtain
\begin{eqnarray}
D^s(\boldsymbol{\lambda}^{Rs})< D(\boldsymbol{\lambda}^{R*})+ \varepsilon'
\end{eqnarray}
which is the upper bound given in Proposition \ref{P:converg}-(ii).

To establish the lower bound, define first the average weighted power consumption as
\begin{equation}\label{E:power_consumption}
\bar{P}(\mathbf{R}(\mathbf{J}),\mathbf{W}(\mathbf{J})):=\sum_{\forall \mathbf{J}}\sum_{m=1}^{M}[\boldsymbol{\mu}]_m\sum_{k=1}^{K}\Upsilon_{\mathcal{R}([\mathbf{J}]_{m,k})}([{\mathbf{R(\mathbf{J})}}]_{m,k})
[\mathbf{W(\mathbf{J})}]_{m,k}\Pr\{\mathbf{J}\}.
\end{equation}
Since the problem in \eqref{E:optRA} has zero duality gap, the optimum primal and dual values coincide; hence
\begin{equation}\label{E:eq_prim_dual} \bar{P}^*=\bar{P}(\mathbf{R}^*(\mathbf{J},\boldsymbol{\lambda}^{R*}),\mathbf{W}^*(\mathbf{J},\boldsymbol{\lambda}^{R*}))=D(\boldsymbol{\lambda}^{R*}).
\end{equation}
On the other hand, it holds that
\begin{equation}\label{E:smooth_dual_as_power}
\bar{P}(\mathbf{R}^*(\mathbf{J},\boldsymbol{\lambda}^{Rs}),\mathbf{W}^s(\mathbf{J},\boldsymbol{\lambda}^{Rs}))=D^s(\boldsymbol{\lambda}^{Rs}). \end{equation}
This is because the iterations in Proposition \ref{P:converg}-(i) only converge when $\partial^s D(\boldsymbol{\lambda}^{Rs})=\mathbf{0}$; the smooth subgradient being zero requires all the average rate constraints to be satisfied with equality; and the latter implies that the only remaining term in the Lagrangian is $\bar{P}(\mathbf{R}^*(\mathbf{J},\boldsymbol{\lambda}^{Rs}),\mathbf{W}^s(\mathbf{J},\boldsymbol{\lambda}^{Rs}))$; cf. \eqref{E:power_consumption}, \eqref{E:simpl_Lagrangian}, and the definition of $D^s(\boldsymbol{\lambda}^{Rs})$ in Lemma \ref{L:propDSmoth}. Finally, since $\mathbf{R}^*(\mathbf{J},\boldsymbol{\lambda}^{Rs})$ and $\mathbf{W}^s(\mathbf{J},\boldsymbol{\lambda}^{Rs})$ are feasible primal variables, it holds that $\bar{P}^*\leq\bar{P}(\mathbf{R}^*(\mathbf{J},\boldsymbol{\lambda}^{Rs}),\mathbf{W}^s(\mathbf{J},\boldsymbol{\lambda}^{Rs}))$. Using \eqref{E:eq_prim_dual} and \eqref{E:smooth_dual_as_power}, the latter inequality yields $D(\boldsymbol{\lambda}^{R*})\leq D^s(\boldsymbol{\lambda}^{Rs})$, which corresponds to the lower bound given in Proposition \ref{P:converg}-(ii).

At this point, it is worth clarifying a potentially misleading implication of Proposition \ref{P:converg}. Once the exact value of $\boldsymbol{\lambda}^{Rs}$ is found after using iterations in \eqref{E:off-line-iter}, one can use Lemma \ref{L:propDSmoth}-(i) to show that $D(\boldsymbol{\lambda}^{Rs})\leq D^s(\boldsymbol{\lambda}^{Rs})$. This implies that the power cost of \emph{the Lagrangian} in \eqref{E:Lagrangian} with primal variables $\mathbf{R}^*(\mathbf{J},\boldsymbol{\lambda}^{Rs})$ and $\mathbf{W}^*(\mathbf{J},\boldsymbol{\lambda}^{Rs})$ used as final solution will be lower than that with the smooth $\mathbf{R}^*(\mathbf{J},\boldsymbol{\lambda}^R)$ and $\mathbf{W}^s(\mathbf{J},\boldsymbol{\lambda}^R)$. Nevertheless, $\mathbf{R}^*(\mathbf{J},\boldsymbol{\lambda}^{Rs})$ and $\mathbf{W}^*(\mathbf{J},\boldsymbol{\lambda}^{Rs})$ cannot be used as a better approximation to the optimal solution $\mathbf{R}^*(\mathbf{J},\boldsymbol{\lambda}^{R*})$ and $\mathbf{W}^*(\mathbf{J},\boldsymbol{\lambda}^{R*})$ because $\mathbf{R}^*(\mathbf{J},\boldsymbol{\lambda}^{Rs})$ and $\mathbf{W}^*(\mathbf{J},\boldsymbol{\lambda}^{Rs})$ may (and most likely will) fail to satisfy the average rate constraints in \eqref{E:optRA}, leading to infeasibility from a primal point of view. On the other hand, the primal variables $\mathbf{R}^*(\mathbf{J},\boldsymbol{\lambda}^R)$ and $\mathbf{W}^s(\mathbf{J},\boldsymbol{\lambda}^R)$ give rise to a slightly higher dual objective (thus higher power cost in the Lagrangian), but they are guaranteed to be feasible and tightly satisfy the average rate constraints.

\section*{Appendix F:  Proof of Proposition \ref{P:cq4}}

Using (\ref{E:rateopt_nom}) and (\ref{E:phi_chann_ind}) we can write
$[\mathbf{C}_W]_{m,k}:=$$\Upsilon_{\mathcal{R}_{m,k}(\mathbf{J})}([\mathbf{R^{*}}]_{m,k})$
$-\dot{\Upsilon}_{\mathcal{R}([\mathbf{J}]_{m,k})}([\mathbf{R^{*}}]_{m,k})$
$[\mathbf{R^{*}}]_{m,k}$. On the one hand, the convexity of $\Upsilon$ guarantees:
${\partial [\mathbf{C}_W]_{m,k}}/{\partial [\mathbf{R}]_{m,k}}=$
$-\ddot{\Upsilon}_{\mathcal{R}([\mathbf{J}]_{m,k})}([\mathbf{R^{*}}]_{m,k})$
$[\mathbf{R^{*}}]_{m,k}$$<0$; on the other hand, it is assumed that $[\mathbf{R}(j_{m,k}+1)]_{m,k}$$>[\mathbf{R}(j_{m,k})]_{m,k}$. The combination of these two conditions implies that
$[\mathbf{C}_W(j_{m,k}+1)]_{m,k}<[\mathbf{C}_W(j_{m,k})]_{m,k}$, which proves \emph{(i)}. Based on this monotonicity property, we prove next \emph{(ii)} and \emph{(iii)}.

If a vector $\mathbf{j'}$ belongs to the set in \emph{(ii)}, then
$[\mathbf{C}_W([\mathbf{j'}]_{m'})]_{m',k}\geq[\mathbf{C}_W([\mathbf{j}]_{m'})]_{m',k}\geq
[\mathbf{C}_W([\mathbf{j}]_m)]_{m,k}=[\mathbf{C}_W([\mathbf{j'}]_{m})]_{m,k}~\forall
m'$, and therefore \emph{(ii)} follows. Observe that the first inequality is
due the condition $[\mathbf{j}']_{m'}\leq[\mathbf{j}]_m'~\forall m'$
in \emph{(ii)} and the decreasing behavior of
$[\mathbf{C}_W(j_{m,k}+1)]_{m,k}$. The second holds because $m \in \mathcal{M}(\mathbf{j},k)$ but $m' \notin \mathcal{M}(\mathbf{j},k)$, and the third is due to the condition
$[\mathbf{j}']_{m}=[\mathbf{j}]_m$ in \emph{(ii)}.

If a vector $\mathbf{j'}$ belongs to the set in \emph{(iii)}, since $[\mathbf{j}']_{m'}\geq[\mathbf{j}]_{m'}$, then
$[\mathbf{C}_W([\mathbf{j'}]_{m'})]_{m',k}\leq[\mathbf{C}_W([\mathbf{j}]_{m'})]_{m',k}$
(better the channel, lower the cost), and therefore $\min \{ [\mathbf{C}_W(\mathbf{j'})]_{k}\}\leq\min
\{[\mathbf{C}_W(\mathbf{j})]_{k}\}$. Furthermore, since $\mathbf{j}\notin
\mathcal{J}_k^{m,l}$, it holds that $\min
\{[\mathbf{C}_W(\mathbf{j})]_{k}\}<[\mathbf{C}_W([\mathbf{j}]_m)]_{m,k}$. On the other hand,
using that $[\mathbf{j}']_{m}=[\mathbf{j}]_{m}$, it follows that
$[\mathbf{C}_W([\mathbf{j}]_m)]_{m,k}=[\mathbf{C}_W([\mathbf{j'}]_m)]_{m,k}$.
Based on these observations it is inferred that $\min
\{[\mathbf{C}_W(\mathbf{j'})]_{k}\}<[\mathbf{C}_W([\mathbf{j'}]_m)]_{m,k}$,
which proves \emph{(iii)}.

\section*{Appendix G:  Proof of Convexity of Eqs. \eqref{E:p-r_ErgCap}, \eqref{E:up_BER_max} and
\eqref{E:up_BER_av}}

To show the convexity of \eqref{E:p-r_ErgCap}, recall that if $x=f^{-1}(y)$ is the inverse function of $y=f(x)$, then $\dot{f^{-1}}(y)=1/(\dot{f}[f^{-1}(y)])$. Using the chain rule of differentiation it follows that $\ddot{f^{-1}}(y)=-\ddot{f}[f^{-1}(y)]/$ $\left(\dot{f}[f^{-1}(y)]\right)^3$. Substituting $f=\Upsilon^{-1}$ and $f^{-1}=\Upsilon$ into the last equality yields
\begin{equation}\label{E:second_der_erg_capacity}
\ddot{\Upsilon}(x)=\frac{-\ddot{\Upsilon^{-1}}[\Upsilon(x)]}{\left(\dot{\Upsilon^{-1}}[\Upsilon(x)]\right)^3}.
\end{equation}
By the definition of $\Upsilon^{-1}$ in \eqref{E:r-p_ErgCap}, it can be readily checked that $\dot{\Upsilon^{-1}}>0$ and $\ddot{\Upsilon^{-1}}<0$. These inequalities imply that \eqref{E:second_der_erg_capacity} is positive, and hence $\Upsilon$ is strictly convex.

The convexity of (\ref{E:up_BER_max}) is straightforward by readily confirming positivity of
\begin{equation}\label{E:der_up_BER_max}
\ddot{\Upsilon}_{\mathcal{R}([\mathbf{J}]_{m,k})}\left(x\right)=\frac{2^{x}\ln(4)
\ln(\kappa_1/\epsilon_{\max})}{\kappa_2q_{m,k,[\mathbf{J}]_{m,k}-1}}.
\end{equation}

Finally, to show the convexity of (\ref{E:up_BER_av}), define first
\begin{eqnarray}\label{E:f_ep_aux}
f_{\epsilon}(x,y):=~\frac{\overline{\epsilon}}{\kappa_1}\int_{q_{m,k,[\mathbf{j}]_m-1}}^{q_{m,k,[\mathbf{j}]_m}}
e^{-\frac{g_{m,k}}{\overline{g}_{m,k}}}dg_{m,k}
-\int_{q_{m,k,[\mathbf{j}]_m-1}}^{q_{m,k,[\mathbf{j}]_m}}
e^{-\frac{g_{m,k}}{\overline{g}_{m,k}}\left(1+\frac{y\overline{g}_{m,k}\kappa_2}{2^x-1}\right)}
dg_{m,k},
\end{eqnarray}
and re-write  $\Upsilon_{\mathcal{R}([\mathbf{J}]_{m,k})}$ as
\begin{equation}\label{E:up_BER_av_via_f_ep}
\Upsilon_{\mathcal{R}([\mathbf{J}]_{m,k})}=\left\{\vphantom{\frac{\frac{1^^1}{1^^1}}{\frac{1^^1}{1^^1}}}
x\rightarrow y: f_{\epsilon}(x,y)=0 \right\},
\end{equation}
where $y$ is uniquely determined by the equation
$f_{\epsilon}(x,y)=0$. Since $df_{\epsilon} = \frac{\partial
f_{\epsilon}}{\partial x} d x + \frac{\partial
f_{\epsilon}}{\partial y} \frac{\partial y}{\partial x} d x =0$, and $\frac{\partial y}{\partial x}=\frac{-\partial
f_{\epsilon}/\partial x}{\partial f_{\epsilon}/\partial y}$, substituting
from \eqref{E:f_ep_aux} yields
\begin{eqnarray}
\frac{
\partial y}{\partial x}=\frac{-\partial f_{\epsilon}/\partial
x}{\partial f_{\epsilon}/\partial y}= \frac{
\int_{q_{m,k,[\mathbf{j}]_m-1}}^{q_{m,k,[\mathbf{j}]_m}}
\frac{y2^x\ln(2)\kappa_2}{(2^x-1)^2}g_{m,k}e^{-\frac{g_{m,k}}{\overline{g}_{m,k}}\left(1+\frac{y\overline{g}_{m,k}\kappa_2}{2^x-1}\right)}
dg_{m,k} }
{\int_{q_{m,k,[\mathbf{j}]_m-1}}^{q_{m,k,[\mathbf{j}]_m}}
\frac{\kappa_2}{2^x-1}g_{m,k}e^{-\frac{g_{m,k}}{\overline{g}_{m,k}}\left(1+\frac{y\overline{g}_{m,k}\kappa_2}{2^x-1}\right)}
dg_{m,k} } =\frac{y2^x\ln(2)}{2^x-1}
\end{eqnarray}
and for the second derivative
\begin{eqnarray}\label{E:second_der_upsilo_av_BER}
\frac{\partial^2 y}{\partial x^2}=\frac{\partial y}{\partial
x}\frac{2^x}{2^x-1}+y\frac{-2^x\ln(2)}{(2^x-1)^2}=\frac{y2^x\ln(2)}{2^x-1}.
\end{eqnarray}

Since $x$ and $y$ (rate and power) are positive, it follows readily that
${\partial^2 y}/{\partial x^2}>0$.

\newpage

\bibliography{journal_arx_amggjr09}

\end{document}